%% file: SCFrame.tex
\documentclass[conference]{IEEEtran}
\IEEEoverridecommandlockouts
\usepackage{cite}
\usepackage{amsmath,amssymb,amsfonts}
\usepackage{graphicx}
\usepackage{textcomp}
\usepackage{xcolor}

\usepackage{microtype}
\usepackage{booktabs} 
\usepackage{xspace}
\usepackage{color}
\usepackage{enumitem}
\usepackage{times}
\usepackage{algcompatible}
\usepackage{algorithm}
\usepackage[]{algpseudocode}
\algrenewcommand\alglinenumber[1]{\tiny #1:}
\usepackage{listings}
\usepackage{verbatim}
\usepackage{float}
\usepackage{pifont}
\usepackage{lineno}
\usepackage{float}
\usepackage{array}
\usepackage{soul}
\usepackage{hyperref}
\usepackage{hyphenat}
\usepackage{setspace}
\usepackage{todonotes}
\usepackage{multirow}

\algnewcommand\algorithmicswitch{\textbf{switch}}
\algnewcommand\algorithmiccase{\textbf{case}}
\algnewcommand\algorithmicassert{\texttt{assert}}
\algnewcommand\Assert[1]{\State \algorithmicassert(#1)}%
\algdef{SE}[SWITCH]{Switch}{EndSwitch}[1]{\algorithmicswitch\ #1\ \algorithmicdo}{\algorithmicend\ \algorithmicswitch}%
\algdef{SE}[CASE]{Case}{EndCase}[1]{\algorithmiccase\ #1}{\algorithmicend\ \algorithmiccase}%
\algtext*{EndSwitch}%
\algtext*{EndCase}%

\makeatletter
\newcommand{\printfnsymbol}[1]{%
	\textsuperscript{\@fnsymbol{#1}}%
}
\makeatother

\input{macros}

\def\BibTeX{{\rm B\kern-.05em{\sc i\kern-.025em b}\kern-.08em
    T\kern-.1667em\lower.7ex\hbox{E}\kern-.125emX}}
\begin{document}

\title{ An Efficient Framework for Optimistic Concurrent Execution of Smart Contracts\printfnsymbol{1}\printfnsymbol{2}\thanks{\scriptsize{\printfnsymbol{1}Author sequence follows lexical order of last names.\newline
			\printfnsymbol{2} This paper has been accepted as full paper in PDP'19.\newline\printfnsymbol{3}The proposal of this paper has been accepted in Doctoral Symposium, ICDCN 2019.	
}}
}
\author{Parwat Singh Anjana, Sweta Kumari, Sathya Peri, Sachin Rathor, and Archit Somani\\
	Department of Computer Science and Engineering, IIT Hyderabad, India\\
	(cs17resch11004, cs15resch01004, sathya\_p, cs18mtech01002, cs15resch01001)@iith.ac.in }

\maketitle

\begin{abstract}
Blockchain platforms such as Ethereum and several others execute \emph{complex transactions} in blocks through user-defined scripts known as \emph{smart contracts}. Normally, a block of the chain consists of multiple transactions of \SContract{s} which are added by a \emph{miner}. To append a correct block into the blockchain, miners execute these transactions of \SContract{s} sequentially. Later the \emph{validators} serially re-execute the \scontract transactions of the block. If the validators agree with the final state of the block as recorded by the miner, then the block is said to be validated. It is then added to the blockchain using a consensus protocol. In Ethereum and other blockchains that support cryptocurrencies, a miner gets an incentive every time such a valid block successfully added to the \bc{}. 


In most of the current day blockchains the miners and validators execute the \scontract transactions serially. In the current era of multi-core processors, by employing the serial execution of the transactions, the miners and validators fail to utilize the cores properly and as a result, have poor throughput. By adding concurrency to \SContract{s} execution, we can achieve better efficiency and higher throughput. In this paper, we develop an efficient framework to execute the \SContract{} transactions concurrently using optimistic \emph{Software Transactional Memory systems} (STMs).

Miners execute \SContract{} transactions concurrently using multi-threading to generate the final state of blockchain. STM is used to take care of synchronization issues among the transactions and ensure atomicity. Now when the validators also execute the transactions (as a part of validation) concurrently using multi-threading, then the validators may get a different final state depending on the order of execution of conflicting transactions. To avoid this, the miners also generate a block graph of the transactions during the concurrent execution and store it in the block. This graph captures the conflict relations among the transactions and is generated concurrently as the transactions are executed by different threads. 

The miner proposes a block which consists of set of transactions, block graph, hash of the previous block, and final state of each shared data-objects. 
Later, the validators re-execute the same \SContract{} transactions concurrently and deterministically with the help of block graph given by the miner to verify the final state. If the validation is successful then proposed block appended into the blockchain and miner gets incentive otherwise discard the proposed block. 

We execute the smart contract transactions concurrently using Basic Time stamp Ordering (BTO) and Multi-Version Time stamp Ordering (MVTO) protocols as optimistic STMs. BTO and MVTO miner achieves 3.6x and 3.7x average speedups over serial miner respectively. Along with, BTO and MVTO validator outperform average 40.8x and 47.1x than serial validator respectively.  

\end{abstract}

\noindent

\input{intro}

\input{Background}
\input{objective}

\input{pm}

\input{results}

\input{conclusion}
\vspace{-.2cm}
\bibliography{citations}
\clearpage
\input{appendix}

\end{document}

%% file: macros.tex
\newcommand{\punt}[1]{}
\newcommand{\cmnt}[1]{}

\newtheorem{theorem}{Theorem}
\newtheorem{lemma}[theorem]{Lemma}

\newtheorem{requirement}[theorem]{Requirement}
\newenvironment{proof}[1][Proof]{\noindent\textbf{#1.} }{} 

\newcommand{\secref}[1]{Section~\ref{sec:#1}}
\newcommand{\figref}[1]{Figure~\ref{fig:#1}}

\newcommand{\algoref}[1]{{Algorithm \ref{alg:#1}}}
\newcommand{\subsecref}[1]{SubSection~\ref{subsec:#1}}

\newcommand{\apnref}[1]{Appendix~\ref{apn:#1}}

\newcommand{\Lineref}[1]{Line~\ref{lin:#1}}

\newcommand{\remove}[1]{}

\newcommand{\ignore}[1]{}

\newcommand{\op} {operation}

\newcommand{\comm}[1] {\textit{committed(#1)}}
\newcommand{\aborted}[1] {\textit{aborted}(#1)}
\newcommand{\txns}[1] {\textit{txns}(#1)}

\newcommand{\evts}[1] {evts(#1)}
\newcommand{\ssch} {sub-history\xspace}

\newcommand{\valid} {valid}

\newcommand{\legal} {legal}

\newcommand{\tobj} {\emph{t-object}}

\newcommand{\opq} {opaque\xspace}

\newcommand{\opty} {opacity\xspace}

\newcommand{\sble} {serializable\xspace}
\newcommand{\sbty} {serializability\xspace}

\newcommand{\tseq} {t-sequential\xspace}

\newcommand{\mv} {mv}
\newcommand{\rf} {rf}
\newcommand{\rt} {rt}
\newcommand{\mvto} {MVTO\xspace}

\newcommand{\cc} {correctness-criterion\xspace}

\newcommand{\readi} {\emph{$read_i$}\xspace}
\newcommand{\cl} {\emph{confList}\xspace}

\newcommand{\tryc} {STM.tryC\xspace}

\newcommand{\trya} {STM.tryA}

\newcommand{\tcntr} {\emph{tCounter}\xspace}
\newcommand{\shl} {\emph{shl}\xspace}
\newcommand{\mr} {\emph{max\_r}\xspace}
\newcommand{\mw} {\emph{max\_w}\xspace}
\newcommand{\cminer} {Concurrent Miner\xspace}
\newcommand{\cvalidator} {Concurrent Validator\xspace}
\newcommand{\conminer} {\emph{concurrent miner}\xspace}
\newcommand{\convalidator} {\emph{concurrent validator}\xspace}

\newcommand{\CG} {\emph{Conflict Graph}\xspace}
\newcommand{\cgraph} {\emph{conflict graph}\xspace}
\newcommand{\Cgraph} {\emph{Conflict graph}\xspace}
\newcommand{\bg} {block graph\xspace}
\newcommand{\aul} {\emph{auList[]}\xspace}
\newcommand{\confg} {\emph{confGraph}\xspace}
\newcommand{\gconfl} {\emph{getConfList}\xspace}
\newcommand{\vp} {\emph{vPred}\xspace}
\newcommand{\vc} {\emph{vCurr}\xspace}
\newcommand{\vn} {\emph{vNext}\xspace}
\newcommand{\ep} {\emph{ePred}\xspace}
\newcommand{\ec} {\emph{eCurr}\xspace}
\newcommand{\en} {\emph{eNext}\xspace}
\newcommand{\vl} {\emph{vList}\xspace}
\newcommand{\el} {\emph{eList}\xspace}
\newcommand{\inc} {\emph{inCnt}\xspace}
\newcommand{\txlog} {\emph{txlog}\xspace}
\newcommand{\adde} {\emph{addEdge}\xspace}
\newcommand{\addv} {\emph{addVert}\xspace}
\newcommand{\egn} {\emph{eNode}\xspace}
\newcommand{\vgn} {\emph{vNode}\xspace}
\newcommand{\enode} {\emph{eNode}\xspace}
\newcommand{\vnode} {\emph{vNode}\xspace}
\newcommand{\searchl} {\emph{searchLocal}\xspace}
\newcommand{\searchg} {\emph{searchGlobal}\xspace}
\newcommand{\bc} {blockchain\xspace}
\newcommand{\BC} {Blockchain\xspace}
\newcommand{\scontract} {smart contract\xspace}
\newcommand{\SContract} {smart contract\xspace}
\newcommand{\Scontract} {Smart contract\xspace}
\newcommand{\au} {atomic-unit\xspace}
\newcommand{\miner} {miner\xspace}

\newcommand{\Miner} {miner\xspace}
\newcommand{\Validator} {validator\xspace}
\newcommand{\exec} {\emph{executeCode}\xspace}
\newcommand{\hb} {happens-before\xspace}

\newcommand{\nc} {\emph{nCount}\xspace}

\newcommand{\vt} {\emph{vTail}\xspace}
\newcommand{\vh} {\emph{vHead}\xspace}
\newcommand{\eh} {\emph{eHead}\xspace}
\newcommand{\et} {\emph{eTail}\xspace}
\newcommand{\tl} {\emph{thLog}\xspace}
\newcommand{\cachel} {\emph{cacheList}\xspace}

\newcommand{\vtuple} {\emph{v\_tuple}\xspace}
\newcommand{\dtuple} {\emph{d\_tuple}\xspace}
\newcommand{\ctuple} {\emph{c\_tuple}\xspace}
\newcommand{\livel} {liveList\xspace}
\newcommand{\begtrans} {\emph{begin}\xspace}
\newcommand{\checkv} {check\_versions\xspace}
\newcommand{\init} {\emph{initialize}\xspace}
\newcommand{\find} {find\_lts\xspace}

\newcommand{\lastw} {lastWrite}

\newcommand{\mvve} {MVVE\xspace}
\newcommand{\vie} {VE\xspace}
\newcommand{\ce} {CE\xspace}

\newcommand{\rs}{rset\xspace}
\newcommand{\ws}{wset\xspace}

\newcommand{\begt} {STM.begin\xspace}
\newcommand{\tread} {STM.read\xspace}
\newcommand{\twrite} {STM.write\xspace}

\newcommand{\commit}{\mathcal{C}}
\newcommand{\abort}{\mathcal{A}}
\newcommand{\mth} {method\xspace}

\newcommand {\incomp}[1] {#1.incomp}
\newcommand {\live}[1] {#1.live}

\newcommand{\shist}[2] {#2.subhist(#1)\xspace}

%% file: intro.tex
\section{Introduction}
\label{sec:intro}
It is commonly believed that \bc{} is a revolutionary technology for doing business over the Internet. \BC is a decentralized, distributed database or ledger of records. Cryptocurrencies such as Bitcoin \cite{Nakamoto:Bitcoin:2009} and Ethereum \cite{ethereum} were the first to popularize the \bc technology. \BC{s} ensure that the records are tamper-proof but publicly readable. 

Basically, the \bc network consists of multiple peers (or nodes) where the peers do not necessarily trust each other. Each node maintains a copy of the distributed ledger. \emph{Clients}, users of the \bc, send requests or \emph{transactions} to the nodes of the \bc called as \emph{miners}. The miners collect multiple transactions from the clients and form a \emph{block}. Miners then propose these blocks to be added to the \bc. 
They follow a global consensus protocol to agree on which blocks are chosen to be added and in what order. While adding a block to the \bc, the miner incorporates the hash of the previous block into the current block. This makes it difficult to tamper with the distributed ledger. The resulting structure is in the form of a linked list or a chain of blocks and hence the name \bc. 

The transactions sent by clients to miners are part of a larger code called as \emph{\scontract{s}} that provide several complex services such as managing the system state, ensuring rules, or credentials checking of the parties involved \cite{Dickerson+:ACSC:PODC:2017}. \Scontract{s} are like a `class' in programming languages that encapsulate data and methods which operate on the data. The data represents the state of the \scontract (as well as the \bc) and the \mth{s} (or functions) are the transactions that possibly can change contract state. A transaction invoked by a client is typically such a \mth or a collection of \mth{s} of the \scontract{s}. Ethereum uses Solidity \cite{Solidity} while Hyperledger supports language such as Java, Golang, Node.js etc. 

\vspace{1mm}
\noindent
\textbf{Motivation for Concurrent Execution of Smart Contracts: }
As observed by Dickerson et al. \cite{Dickerson+:ACSC:PODC:2017}, \scontract transactions are executed in two different contexts specifically in Ethereum. First, they are executed by miners while forming a block. A miner selects a sequence of client request transactions, executes the smart contract code of these transactions in sequence, transforming the state of the associated contract in this process. The miner then stores the sequence of transactions, the resulting final state of the contracts in the block along with the hash of the previous block. After creating the block, the miner proposes it to be added to the \bc through the consensus protocol.

Once a block is added, the other peers in the system, referred to as \emph{validators} in this context, validate the contents of the block. They re-execute the \scontract transactions in the block to verify the block's final states match or not. If final states match, then the block is accepted as valid and the miner who appended this block is rewarded. Otherwise, the block is discarded. Thus the transactions are executed by every peer in the system. In this setting, it turns out that the validation code runs several times more than miner code \cite{Dickerson+:ACSC:PODC:2017}. 

This design of \scontract execution is not very efficient as it does not allow any concurrency. Both the miner and the validator execute transactions serially one after another. In today's world of multi-core systems, the serial execution does not utilize all the cores and hence results in lower throughput. This limitation is not specific only to Ethereum but almost all the popular \bc{s}. Higher throughput means more number of transactions executed per unit time by miners and validators which clearly will be desired by both of them.  

\ignore{
\figref{sece} illustrates the motivation behind the execution of smart contracts by concurrent miner over serial miner. Consider \figref{sece} (a) which consists of two transactions $T_1$, and $T_2$ executed by the serial miner. Here, $T_1$, and $T_2$ are writing on data-objects $x$, and $y$ respectively. Due to the  serial execution by miner, all the transactions are executing serially although they are working on different data-objects which tends to limit the throughput of miner. Whereas \figref{sece} (b) represents the concurrent execution by miner with same scenario as \figref{sece} (a) where $T_1$ and $T_2$ are running concurrently because they are working on different data-objects. Hence, concurrent execution by miner improves the throughput as compare to serial miner.
\begin{figure}
	\centerline{\scalebox{0.35}{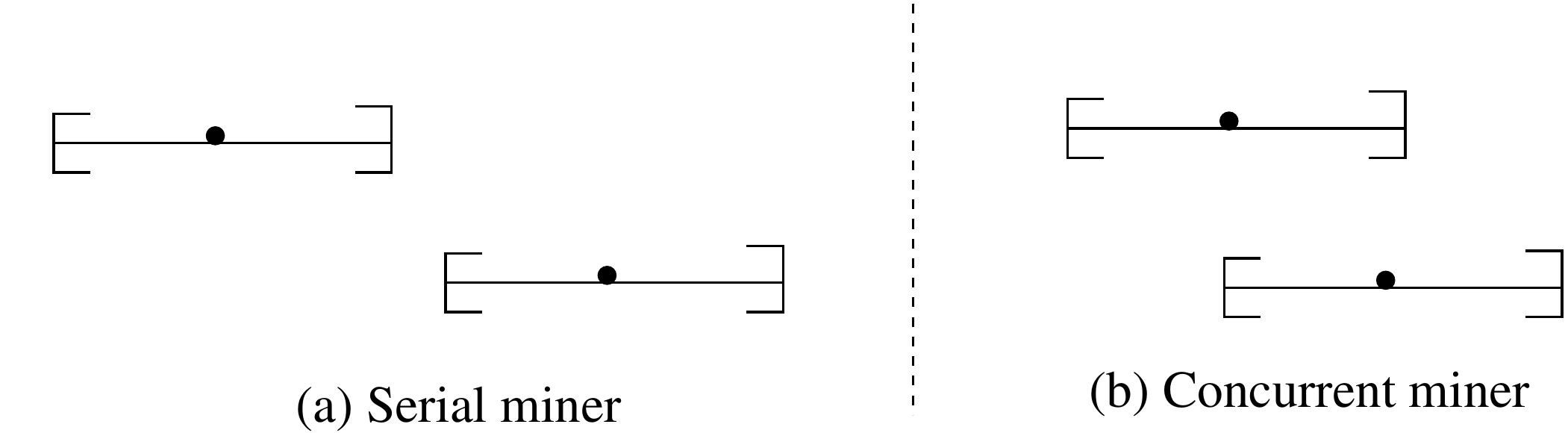}}
	\caption{Efficient execution of smart contracts}
	\label{fig:sece}
\end{figure}
}

But the concurrent execution of \scontract transactions is not an easy task. The various transactions requested by the clients could consist of conflicting access to the shared data-objects. Arbitrary execution of these transactions by the miners might result in the data-races leading to the inconsistent final state of the \bc. Unfortunately, it is not possible to statically identify if two contract transactions are conflicting or not since they are developed in Turing-complete languages. The common solution for correct execution of concurrent transactions is to ensure that the execution is \emph{\sble} \cite{Papad:1979:JACM}. A usual \cc in databases, \sbty ensure that the concurrent execution is equivalent to some serial execution of the same transactions. Thus the miners must ensure that their execution is \sble \cite{Dickerson+:ACSC:PODC:2017} or one of its variants as described later.

\begin{figure}[H]
	\centerline{\scalebox{0.29}{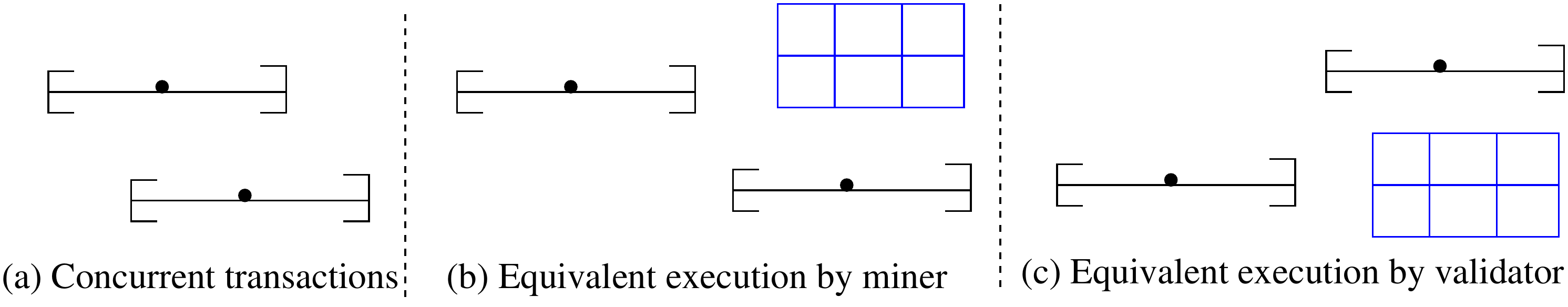}}
\vspace{-.3cm} 	\caption{Execution of concurrent transactions by miner and validator}
	\label{fig:conmv}
\end{figure}

The concurrent execution of the \scontract transactions of a block by the validators although highly desirable can further complicate the situation. Suppose a miner ensures that the concurrent execution of the transactions in a block are \sble. Later a validator executes the same transactions concurrently. But during the concurrent execution, the validator may execute two conflicting transactions in an order different from what was executed by the miner. Thus the serialization order of the miner is different from the validator. Then this can result in the validator obtaining a final state different from what was obtained by the miner. Consequently, the validator may incorrectly reject the block although it is valid. \figref{conmv} illustrates this in the following example. \figref{conmv} (a) consists of two concurrent conflicting transactions $T_1$ and $T_2$ working on same shared data-objects $x$ which are part of a block. \figref{conmv} (b) represents the concurrent execution by miner with an equivalent serial schedule as $T_1$, $T_2$ and final state (or FS) as 20 from the initial state (or IS) 0. Whereas \figref{conmv} (c), shows the concurrent execution by a validator with an equivalent serial schedule as $T_2$, $T_1$, and final state as 10 from IS 0 which is different from the final state proposed by the miner. Such a situation leads to rejection of the valid block by the validator which is undesirable.
\vspace{-.5cm}
\vspace{-.4cm}
These important issues were identified by Dickerson et al. \cite{Dickerson+:ACSC:PODC:2017} who proposed a solution of concurrent execution for both the miners and validators. In their solution, the miners concurrently execute the transactions of a block using abstract locks and inverse logs to generate a serializable execution. Then, to enable correct concurrent execution by the validators, the miners also provide a \emph{\hb} graph in the block. The \hb graph is a direct acyclic graph over all the transaction of the block. If there is a path from a transaction $T_i$ to $T_j$ then the validator has to execute $T_i$ before $T_j$. Transactions with no path between them can execute concurrently. The validator using the \hb graph in the block executes all the transactions concurrently using the fork-join approach. This methodology ensures that the final state of the \bc generated by the miners and the validators are the same for a valid block and hence not rejected by the validators. The presence of tools such as a \hb graph in the block provides greater enhancement to validators to consider such blocks as it helps them execute quickly by means of parallelization as opposed to a block which does not have any tools for parallelization. This, in turn, entices the miners to provide such tools in the block for concurrent execution by the validators. 


\ignore {

\figref{cminer}, illustrates the functionality of concurrent miner which consists of six steps. It has two or more serial miner and one concurrent miner competing to each other to propose a block in blockchain. Whoever will propose a block first that miner have a chance to get strong incentive. So the challenge here is to execute the task of miner concurrently. All the miners are getting the set of transactions from distributed shared memory. As we discussed above, serial miner executes the transactions one after another and propose the block. Where as concurrent miner executes the non-conflicting transactions concurrently with the help of Transactional Memory (TM) and finally propose a block. Complete details about the \figref{cminer} presents in the \subsecref{cminer}.

\begin{figure}
	\centering
	\captionsetup{justification=centering}
	\centerline{\scalebox{0.45}{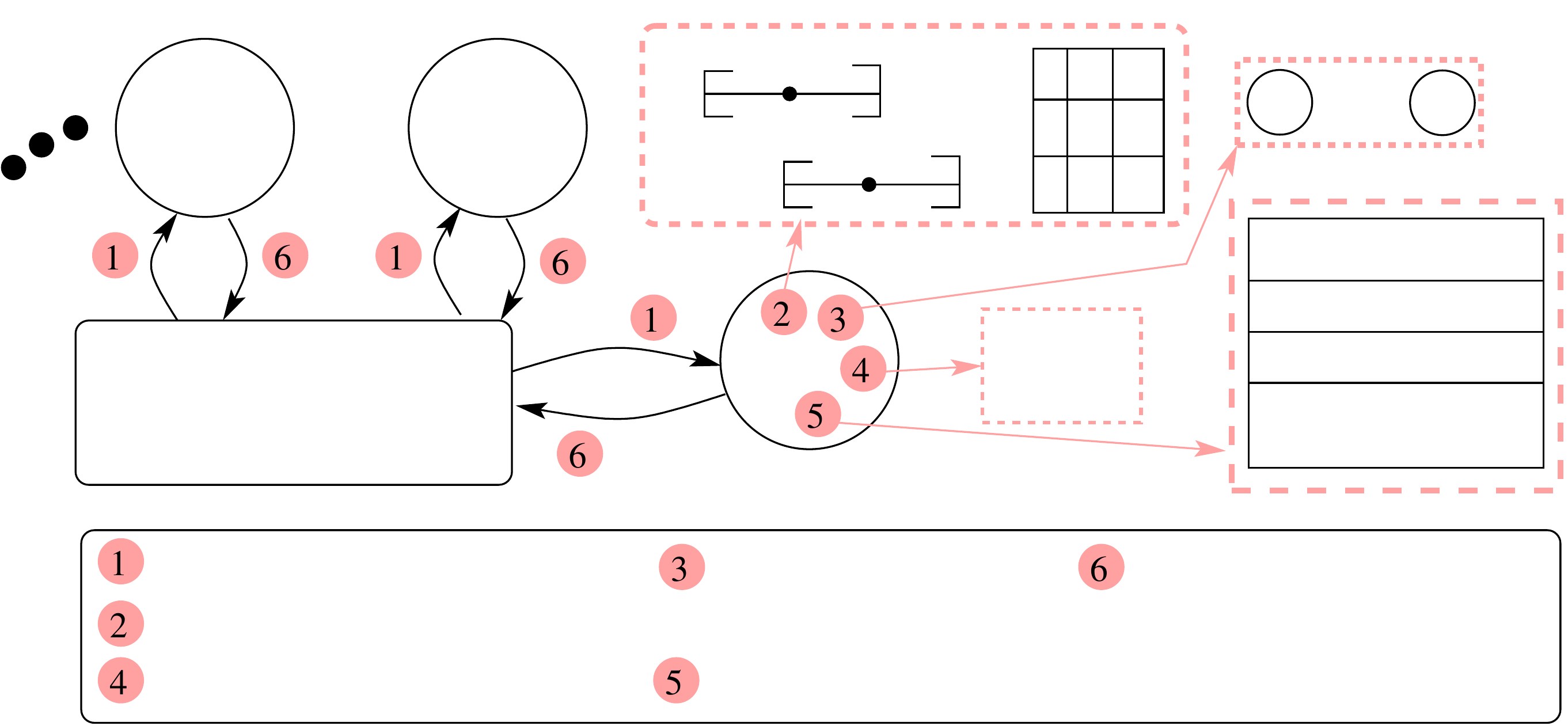}}
	\caption{Execution of Concurrent Miner}
	\label{fig:cminer}
\end{figure}

}

\vspace{1mm}
\noindent
\textbf{Our Solution Approach - Optimistic Concurrent Execution and Lock-Free Graph: } Dickerson et al. \cite{Dickerson+:ACSC:PODC:2017} developed a solution to the problem of concurrent miner and validators using locks and inverse logs. It is well known that locks are pessimistic in nature. So, in this paper, we explore a novel and efficient framework for concurrent miners using optimistic Software Transactional Memory Systems (STMs).

The requirement of the miner, as explained above, is to concurrently execute the \scontract transactions correctly and output a graph capturing dependencies among the transactions of the block such as \hb graph. We denote this graph as \emph{\bg} (or BG). In the proposed solution, the miner uses the services of an optimistic STM system to concurrently execute the \scontract transactions. Since STMs also work with transactions, we differentiate between \scontract transactions and STM transactions. The STM transactions invoked by an STM system is a piece of code that it tries to execute atomically even in presence of other concurrent STM transactions. If the STM system is not able to execute it atomically, then the STM transaction is aborted. 

The expectation of a \scontract transaction is that it will be executed serially. Thus, when it is executed in a concurrent setting, it is expected to be executed atomically (or serialized). But unlike STM transaction, a \scontract transaction cannot be committed or aborted. Thus to differentiate between \scontract transaction from STM transaction, we denote \scontract transaction as \emph{Atomic Unit} or \emph{\au} and STM transaction as transaction in the rest of the document. Thus the miner uses the STM system to invoke a transaction for each \au. In case the transaction gets aborted, then the STM repeatedly invokes new transactions for the same \au until a transaction invocation eventually commits. 

A popular correctness guarantee provided by STM systems is \emph{\opty} \cite{GuerKap:Opacity:PPoPP:2008} which is stronger than \sbty. Opacity like \sbty requires that the concurrent execution including the aborted transactions be equivalent to some serial execution. This ensures that even aborted transaction reads consistent value until the point of abort. As a result, that the application such as a miner using an STM does not encounter any undesirable side-effects such as crash failures, infinite loops, divide by zero etc. STMs provide this guarantee by executing optimistically and support atomic (\opq) reads, writes on \emph{transactional objects} or \emph{\tobj{s}}. 

Among the various STMs available, we have chosen two timestamp based STMs in our design: (1) \emph{Basic Timestamp Ordering} or \emph{BTO} STM \cite[Chap 4]{WeiVoss:TIS:2002:Morg}, maintains only one version for each \tobj. (2) \emph{Multi-Version Timestamp Ordering} or \emph{\mvto} STM \cite{Kumar+:MVTO:ICDCN:2014}, maintains multiple versions corresponding to each \tobj{} which further reduces the number of aborts and improves the throughput. 

The advantage of using timestamp based STM is that in these systems the equivalent serial history is ordered based on the timestamps of the transactions. Thus using the timestamps, the miner can generate the BG of the \au{s}. Dickerson et al. \cite{Dickerson+:ACSC:PODC:2017}, developed the BG in a serial manner. In our approach, the graph is developed by the miner in concurrent and lock-free \cite{HerlihyShavit:Progress:Opodis:2011} manner. 

The validator process creates multiple threads. Each of these threads parses the BG and re-execute the atomic-units for validation. The BG provided by concurrent miner shows dependency among the \au{s}. Each validator thread, claims a node which does not have any dependency, i.e. a node without any incoming edges by marking it. After that, it executes the corresponding \au{s} deterministically. Since the threads execute only those nodes that do not have any incoming edges, the concurrently executing \au{s} will not have any conflicts. Hence the validator threads need not have to worry about synchronization issues. We denote this approach adopted by the validator as a decentralized approach (or Decentralized Validator) as the multiple threads are working on BG concurrently in the absence of master thread. 

The approach adopted by Dickerson et al. \cite{Dickerson+:ACSC:PODC:2017}, works on fork-join in which a master thread allocates different tasks to slave threads. The master thread will identify those \au{s} which do not have any dependencies from the BG and allocates them to different slave threads to work on. In this paper, we compare the performance of both these approaches with the serial validator.

\noindent
\textbf{Contributions of the paper as follows:}
\begin{itemize}
\item Introduce a novel way to execute the \SContract{} transactions by concurrent miner using optimistic STMs.
\item We implement the concurrent miner with the help of BTO and MVTO STM but its generic to any STM protocol.

\item We propose a lock-free graph library to generate the BG.
\item We propose concurrent validator that re-executes the \SContract{} transactions deterministically and efficiently with the help of BG given by concurrent miner.

\item Concurrent miner satisfies correctness criterion as opacity. 
\item We achieve 3.6x and 3.7x average speedups for concurrent miner using BTO and MVTO STM protocol respectively. Along with, BTO and MVTO validator outperform average 40.8x and 47.1x than serial validator respectively.  

\end{itemize}

\noindent
\textbf{Related Work:} The first \emph{blockchain} concept has been given by Satoshi Nakamoto in 2009 \cite{Nakamoto:Bitcoin:2009}. He proposed a system as bitcoin \cite{Nakamoto:Bitcoin:2009} which performs electronic transactions without the involvement of the third party. The term \SContract{} \cite{Nick:PublicN:journals:1997} has been introduced by Nick Szabo. \emph{Smart contract} is an interface to reduce the computational transaction cost and provides secure relationships on public networks. There exist few paper in the literature that works on safety and security concern of smart contracts. Luu et al. \cite{Luu+:DIC:CCS:2015} addresses the waste part of the computational effort by miner that can be utilized and lead to award the incentives.  Delmolino et al.\cite{Delmolino+:SSTCSC:FC:2016} document presents the common pitfall made while designing a secure smart contract. Nowadays, ethereum \cite{ethereum} is one of the most popular smart contract platform which supports a built-in Turing-complete programming language. Ethereum virtual machine (EVM) uses Solidity \cite{Solidity} programming language. Luu et al.\cite{Luu+:MSC:CCS:2016} addresses several security problems and proposed an enhanced mechanism to make the ethereum smart contracts less vulnerable.

Sergey et al. \cite{SergeyandHobor:ACP:2017} elaborates a new perspective between smart contracts and concurrent objects. Zang et al. \cite{ZangandZang:ECSC:WBD:2018} uses any concurrency control mechanism for concurrent miner which delays the read until the corresponding writes to commit and ensures conflict-serializable schedule. Basically, they proposed concurrent validators using MVTO protocol with the help of write sets provided by the concurrent miner. Dickerson et al. \cite{Dickerson+:ACSC:PODC:2017} introduces a speculative way to execute smart contracts by using concurrent miner and concurrent validators. They have used pessimistic software transactional memory systems (STMs) to execute concurrent smart contracts which use rollback if any inconsistency occurs and prove that schedule generated by concurrent miner is \emph{serializable}. We proposed an efficient framework for concurrent execution of the smart contracts using optimistic software transactional memory systems. So, the updates made by a transaction will be visible to shared memory only on commit hence, rollback is not required. Our approach ensures correctness criteria as opacity \cite{GuerKap:Opacity:PPoPP:2008} which considers aborted transactions as well because it read correct values.

Weikum et al. \cite{WeiVoss:TIS:2002:Morg} proposed concurrency control techniques that maintain single-version and multiple versions corresponding to each data-object. STMs \cite{HerlMoss:1993:SigArch,ShavTou:1995:PODC} are alternative to locks for addressing synchronization and concurrency issues in multi-core systems. STMs are suitable for the concurrent executions of smart contracts without worrying about consistency issues. Single-version STMs protocol store single version corresponding to each data-object as BTO STM. It identifies the conflicts between two transactions at run-time and abort one of them and retry again for the aborted transaction. Kumar et al. \cite{Kumar+:MVTO:ICDCN:2014} observe that storing multiple versions corresponding to each data-object reduces the number of aborts and provides greater concurrency that leads to improving the throughput.

%% file: figs/sece.pdf_t
\begin{picture}(0,0)%
\includegraphics{figs/sece.pdf}%
\end{picture}%
\setlength{\unitlength}{4144sp}%
\begingroup\makeatletter\ifx\SetFigFont\undefined%
\gdef\SetFigFont#1#2#3#4#5{%
  \reset@font\fontsize{#1}{#2pt}%
  \fontfamily{#3}\fontseries{#4}\fontshape{#5}%
  \selectfont}%
\fi\endgroup%
\begin{picture}(9592,2651)(2326,-3545)
\put(9406,-1411){\makebox(0,0)[lb]{\smash{{\SetFigFont{17}{20.4}{\rmdefault}{\mddefault}{\updefault}{\color[rgb]{0,0,0}$w(x, 10)$}%
}}}}
\put(8551,-1546){\makebox(0,0)[lb]{\smash{{\SetFigFont{17}{20.4}{\rmdefault}{\mddefault}{\updefault}{\color[rgb]{0,0,0}$T_1$}%
}}}}
\put(10891,-1366){\makebox(0,0)[lb]{\smash{{\SetFigFont{17}{20.4}{\rmdefault}{\mddefault}{\updefault}{\color[rgb]{0,0,0}$C_1$}%
}}}}
\put(10366,-2386){\makebox(0,0)[lb]{\smash{{\SetFigFont{17}{20.4}{\rmdefault}{\mddefault}{\updefault}{\color[rgb]{0,0,0}$w(y, 20)$}%
}}}}
\put(9511,-2521){\makebox(0,0)[lb]{\smash{{\SetFigFont{17}{20.4}{\rmdefault}{\mddefault}{\updefault}{\color[rgb]{0,0,0}$T_2$}%
}}}}
\put(11851,-2341){\makebox(0,0)[lb]{\smash{{\SetFigFont{17}{20.4}{\rmdefault}{\mddefault}{\updefault}{\color[rgb]{0,0,0}$C_2$}%
}}}}
\put(2341,-1636){\makebox(0,0)[lb]{\smash{{\SetFigFont{17}{20.4}{\rmdefault}{\mddefault}{\updefault}{\color[rgb]{0,0,0}$T_1$}%
}}}}
\put(4681,-1456){\makebox(0,0)[lb]{\smash{{\SetFigFont{17}{20.4}{\rmdefault}{\mddefault}{\updefault}{\color[rgb]{0,0,0}$C_1$}%
}}}}
\put(4741,-2491){\makebox(0,0)[lb]{\smash{{\SetFigFont{17}{20.4}{\rmdefault}{\mddefault}{\updefault}{\color[rgb]{0,0,0}$T_2$}%
}}}}
\put(7081,-2311){\makebox(0,0)[lb]{\smash{{\SetFigFont{17}{20.4}{\rmdefault}{\mddefault}{\updefault}{\color[rgb]{0,0,0}$C_2$}%
}}}}
\put(3196,-1501){\makebox(0,0)[lb]{\smash{{\SetFigFont{17}{20.4}{\rmdefault}{\mddefault}{\updefault}{\color[rgb]{0,0,0}$w(x, 10)$}%
}}}}
\put(5596,-2356){\makebox(0,0)[lb]{\smash{{\SetFigFont{17}{20.4}{\rmdefault}{\mddefault}{\updefault}{\color[rgb]{0,0,0}$w(y, 20)$}%
}}}}
\end{picture}%

%% file: figs/conMV.pdf_t
\begin{picture}(0,0)%
\includegraphics{figs/conMV.pdf}%
\end{picture}%
\setlength{\unitlength}{4144sp}%
\begingroup\makeatletter\ifx\SetFigFont\undefined%
\gdef\SetFigFont#1#2#3#4#5{%
  \reset@font\fontsize{#1}{#2pt}%
  \fontfamily{#3}\fontseries{#4}\fontshape{#5}%
  \selectfont}%
\fi\endgroup%
\begin{picture}(13627,2654)(2146,-3728)
\put(4591,-1546){\makebox(0,0)[lb]{\smash{{\SetFigFont{20}{24.0}{\rmdefault}{\mddefault}{\updefault}{\color[rgb]{0,0,0}$C_1$}%
}}}}
\put(10036,-1411){\makebox(0,0)[lb]{\smash{{\SetFigFont{17}{20.4}{\rmdefault}{\mddefault}{\updefault}{\color[rgb]{0,0,0}$FS$}%
}}}}
\put(9496,-1411){\makebox(0,0)[lb]{\smash{{\SetFigFont{17}{20.4}{\rmdefault}{\mddefault}{\updefault}{\color[rgb]{0,0,0}$IS$}%
}}}}
\put(9631,-1861){\makebox(0,0)[lb]{\smash{{\SetFigFont{17}{20.4}{\rmdefault}{\mddefault}{\updefault}{\color[rgb]{0,0,0}0}%
}}}}
\put(9001,-1861){\makebox(0,0)[lb]{\smash{{\SetFigFont{17}{20.4}{\rmdefault}{\mddefault}{\updefault}{\color[rgb]{0,0,0}$x$}%
}}}}
\put(10126,-1861){\makebox(0,0)[lb]{\smash{{\SetFigFont{17}{20.4}{\rmdefault}{\mddefault}{\updefault}{\color[rgb]{1,0,0}20}%
}}}}
\put(15211,-2536){\makebox(0,0)[lb]{\smash{{\SetFigFont{17}{20.4}{\rmdefault}{\mddefault}{\updefault}{\color[rgb]{0,0,0}$FS$}%
}}}}
\put(14671,-2536){\makebox(0,0)[lb]{\smash{{\SetFigFont{17}{20.4}{\rmdefault}{\mddefault}{\updefault}{\color[rgb]{0,0,0}$IS$}%
}}}}
\put(14806,-2986){\makebox(0,0)[lb]{\smash{{\SetFigFont{17}{20.4}{\rmdefault}{\mddefault}{\updefault}{\color[rgb]{0,0,0}0}%
}}}}
\put(14176,-2986){\makebox(0,0)[lb]{\smash{{\SetFigFont{17}{20.4}{\rmdefault}{\mddefault}{\updefault}{\color[rgb]{0,0,0}$x$}%
}}}}
\put(15301,-2986){\makebox(0,0)[lb]{\smash{{\SetFigFont{17}{20.4}{\rmdefault}{\mddefault}{\updefault}{\color[rgb]{1,0,0}10}%
}}}}
\put(3106,-1591){\makebox(0,0)[lb]{\smash{{\SetFigFont{20}{24.0}{\rmdefault}{\mddefault}{\updefault}{\color[rgb]{0,0,0}$w(x, 10)$}%
}}}}
\put(3826,-2536){\makebox(0,0)[lb]{\smash{{\SetFigFont{20}{24.0}{\rmdefault}{\mddefault}{\updefault}{\color[rgb]{0,0,0}$w(x, 20)$}%
}}}}
\put(6661,-1591){\makebox(0,0)[lb]{\smash{{\SetFigFont{20}{24.0}{\rmdefault}{\mddefault}{\updefault}{\color[rgb]{0,0,0}$w(x, 10)$}%
}}}}
\put(5806,-1726){\makebox(0,0)[lb]{\smash{{\SetFigFont{20}{24.0}{\rmdefault}{\mddefault}{\updefault}{\color[rgb]{0,0,0}$T_1$}%
}}}}
\put(8146,-1546){\makebox(0,0)[lb]{\smash{{\SetFigFont{20}{24.0}{\rmdefault}{\mddefault}{\updefault}{\color[rgb]{0,0,0}$C_1$}%
}}}}
\put(9061,-2446){\makebox(0,0)[lb]{\smash{{\SetFigFont{20}{24.0}{\rmdefault}{\mddefault}{\updefault}{\color[rgb]{0,0,0}$w(x, 20)$}%
}}}}
\put(11881,-2401){\makebox(0,0)[lb]{\smash{{\SetFigFont{20}{24.0}{\rmdefault}{\mddefault}{\updefault}{\color[rgb]{0,0,0}$w(x, 20)$}%
}}}}
\put(13366,-2356){\makebox(0,0)[lb]{\smash{{\SetFigFont{20}{24.0}{\rmdefault}{\mddefault}{\updefault}{\color[rgb]{0,0,0}$C_2$}%
}}}}
\put(14221,-1411){\makebox(0,0)[lb]{\smash{{\SetFigFont{20}{24.0}{\rmdefault}{\mddefault}{\updefault}{\color[rgb]{0,0,0}$w(x, 10)$}%
}}}}
\put(13366,-1546){\makebox(0,0)[lb]{\smash{{\SetFigFont{20}{24.0}{\rmdefault}{\mddefault}{\updefault}{\color[rgb]{0,0,0}$T_1$}%
}}}}
\put(2881,-2626){\makebox(0,0)[lb]{\smash{{\SetFigFont{20}{24.0}{\rmdefault}{\mddefault}{\updefault}{\color[rgb]{0,0,0}$T_2$}%
}}}}
\put(2161,-1726){\makebox(0,0)[lb]{\smash{{\SetFigFont{20}{24.0}{\rmdefault}{\mddefault}{\updefault}{\color[rgb]{0,0,0}$T_1$}%
}}}}
\put(5176,-2491){\makebox(0,0)[lb]{\smash{{\SetFigFont{20}{24.0}{\rmdefault}{\mddefault}{\updefault}{\color[rgb]{0,0,0}$C_2$}%
}}}}
\put(8101,-2581){\makebox(0,0)[lb]{\smash{{\SetFigFont{20}{24.0}{\rmdefault}{\mddefault}{\updefault}{\color[rgb]{0,0,0}$T_2$}%
}}}}
\put(10351,-2401){\makebox(0,0)[lb]{\smash{{\SetFigFont{20}{24.0}{\rmdefault}{\mddefault}{\updefault}{\color[rgb]{0,0,0}$C_2$}%
}}}}
\put(15481,-1321){\makebox(0,0)[lb]{\smash{{\SetFigFont{20}{24.0}{\rmdefault}{\mddefault}{\updefault}{\color[rgb]{0,0,0}$C_1$}%
}}}}
\put(11026,-2401){\makebox(0,0)[lb]{\smash{{\SetFigFont{20}{24.0}{\rmdefault}{\mddefault}{\updefault}{\color[rgb]{0,0,0}$T_2$}%
}}}}
\end{picture}%

%% file: figs/cminer.pdf_t
\begin{picture}(0,0)%
\includegraphics{figs/cminer.pdf}%
\end{picture}%
\setlength{\unitlength}{4144sp}%
\begingroup\makeatletter\ifx\SetFigFont\undefined%
\gdef\SetFigFont#1#2#3#4#5{%
  \reset@font\fontsize{#1}{#2pt}%
  \fontfamily{#3}\fontseries{#4}\fontshape{#5}%
  \selectfont}%
\fi\endgroup%
\begin{picture}(12434,5761)(1606,-6068)
\put(11611,-1186){\makebox(0,0)[lb]{\smash{{\SetFigFont{17}{20.4}{\rmdefault}{\mddefault}{\updefault}{\color[rgb]{0,0,0}$T_1$}%
}}}}
\put(12871,-1231){\makebox(0,0)[lb]{\smash{{\SetFigFont{17}{20.4}{\rmdefault}{\mddefault}{\updefault}{\color[rgb]{0,0,0}$T_2$}%
}}}}
\put(9451,-3031){\makebox(0,0)[lb]{\smash{{\SetFigFont{17}{20.4}{\rmdefault}{\mddefault}{\updefault}{\color[rgb]{0,0,0}$Compute$}%
}}}}
\put(9451,-3301){\makebox(0,0)[lb]{\smash{{\SetFigFont{17}{20.4}{\rmdefault}{\mddefault}{\updefault}{\color[rgb]{0,0,0}$Previous$}%
}}}}
\put(9676,-3571){\makebox(0,0)[lb]{\smash{{\SetFigFont{17}{20.4}{\rmdefault}{\mddefault}{\updefault}{\color[rgb]{0,0,0}$Hash$}%
}}}}
\put(2296,-3256){\makebox(0,0)[lb]{\smash{{\SetFigFont{20}{24.0}{\rmdefault}{\mddefault}{\updefault}{\color[rgb]{0,0,0}$Distributed$}%
}}}}
\put(3466,-3616){\makebox(0,0)[lb]{\smash{{\SetFigFont{20}{24.0}{\rmdefault}{\mddefault}{\updefault}{\color[rgb]{0,0,0}$Shared$}%
}}}}
\put(4096,-3976){\makebox(0,0)[lb]{\smash{{\SetFigFont{20}{24.0}{\rmdefault}{\mddefault}{\updefault}{\color[rgb]{0,0,0}$Memory$}%
}}}}
\put(7426,-1771){\makebox(0,0)[lb]{\smash{{\SetFigFont{17}{20.4}{\rmdefault}{\mddefault}{\updefault}{\color[rgb]{0,0,0}$T_2$}%
}}}}
\put(6796,-1051){\makebox(0,0)[lb]{\smash{{\SetFigFont{17}{20.4}{\rmdefault}{\mddefault}{\updefault}{\color[rgb]{0,0,0}$T_1$}%
}}}}
\put(8551,-736){\makebox(0,0)[lb]{\smash{{\SetFigFont{17}{20.4}{\rmdefault}{\mddefault}{\updefault}{\color[rgb]{0,0,0}$C_1$}%
}}}}
\put(7651,-826){\makebox(0,0)[lb]{\smash{{\SetFigFont{17}{20.4}{\rmdefault}{\mddefault}{\updefault}{\color[rgb]{0,0,0}$r_1(x)$}%
}}}}
\put(8281,-1546){\makebox(0,0)[lb]{\smash{{\SetFigFont{17}{20.4}{\rmdefault}{\mddefault}{\updefault}{\color[rgb]{0,0,0}$w_2(y)$}%
}}}}
\put(10261,-1411){\makebox(0,0)[lb]{\smash{{\SetFigFont{17}{20.4}{\rmdefault}{\mddefault}{\updefault}{\color[rgb]{0,0,0}0}%
}}}}
\put(10261,-1861){\makebox(0,0)[lb]{\smash{{\SetFigFont{17}{20.4}{\rmdefault}{\mddefault}{\updefault}{\color[rgb]{0,0,0}0}%
}}}}
\put(10621,-1411){\makebox(0,0)[lb]{\smash{{\SetFigFont{17}{20.4}{\rmdefault}{\mddefault}{\updefault}{\color[rgb]{0,0,0}0}%
}}}}
\put(10621,-1861){\makebox(0,0)[lb]{\smash{{\SetFigFont{17}{20.4}{\rmdefault}{\mddefault}{\updefault}{\color[rgb]{0,0,0}2}%
}}}}
\put(9181,-1456){\makebox(0,0)[lb]{\smash{{\SetFigFont{17}{20.4}{\rmdefault}{\mddefault}{\updefault}{\color[rgb]{0,0,0}$C_2$}%
}}}}
\put(9856,-1411){\makebox(0,0)[lb]{\smash{{\SetFigFont{17}{20.4}{\rmdefault}{\mddefault}{\updefault}{\color[rgb]{0,0,0}$x$}%
}}}}
\put(9856,-1816){\makebox(0,0)[lb]{\smash{{\SetFigFont{17}{20.4}{\rmdefault}{\mddefault}{\updefault}{\color[rgb]{0,0,0}$y$}%
}}}}
\put(10441,-961){\makebox(0,0)[lb]{\smash{{\SetFigFont{17}{20.4}{\rmdefault}{\mddefault}{\updefault}{\color[rgb]{0,0,0}$FS$}%
}}}}
\put(10081,-961){\makebox(0,0)[lb]{\smash{{\SetFigFont{17}{20.4}{\rmdefault}{\mddefault}{\updefault}{\color[rgb]{0,0,0}$IS$}%
}}}}
\put(7471,-3346){\makebox(0,0)[lb]{\smash{{\SetFigFont{17}{20.4}{\rmdefault}{\mddefault}{\updefault}{\color[rgb]{0,0,0}\emph{TM}}%
}}}}
\put(7201,-4201){\makebox(0,0)[lb]{\smash{{\SetFigFont{17}{20.4}{\rmdefault}{\mddefault}{\updefault}{\color[rgb]{0,0,0}\textbf{\emph{Concurrent Miner}}}%
}}}}
\put(2521,-466){\makebox(0,0)[lb]{\smash{{\SetFigFont{17}{20.4}{\rmdefault}{\mddefault}{\updefault}{\color[rgb]{0,0,0}\emph{Serial Miner}}%
}}}}
\put(4996,-466){\makebox(0,0)[lb]{\smash{{\SetFigFont{17}{20.4}{\rmdefault}{\mddefault}{\updefault}{\color[rgb]{0,0,0}\emph{Serial Miner}}%
}}}}
\put(11611,-2356){\makebox(0,0)[lb]{\smash{{\SetFigFont{17}{20.4}{\rmdefault}{\mddefault}{\updefault}{\color[rgb]{0,0,0}Set of Transactions}%
}}}}
\put(2836,-5821){\makebox(0,0)[lb]{\smash{{\SetFigFont{17}{20.4}{\rmdefault}{\mddefault}{\updefault}{\color[rgb]{0,0,0}: Compute Hash of Previous Block}%
}}}}
\put(2836,-5371){\makebox(0,0)[lb]{\smash{{\SetFigFont{17}{20.4}{\rmdefault}{\mddefault}{\updefault}{\color[rgb]{0,0,0}: Concurrent Execution of Transactions by \emph{TM} and Compute the Final State (FS) of Shared \tobj{}}%
}}}}
\put(2836,-4876){\makebox(0,0)[lb]{\smash{{\SetFigFont{17}{20.4}{\rmdefault}{\mddefault}{\updefault}{\color[rgb]{0,0,0}: Set of Transactions}%
}}}}
\put(7291,-4876){\makebox(0,0)[lb]{\smash{{\SetFigFont{17}{20.4}{\rmdefault}{\mddefault}{\updefault}{\color[rgb]{0,0,0}: Conflict Graph}%
}}}}
\put(10621,-4876){\makebox(0,0)[lb]{\smash{{\SetFigFont{17}{20.4}{\rmdefault}{\mddefault}{\updefault}{\color[rgb]{0,0,0}: Send the Proposed Block}%
}}}}
\put(7291,-5776){\makebox(0,0)[lb]{\smash{{\SetFigFont{17}{20.4}{\rmdefault}{\mddefault}{\updefault}{\color[rgb]{0,0,0}: Proposed Block by Concurrent Miner}%
}}}}
\put(11611,-2806){\makebox(0,0)[lb]{\smash{{\SetFigFont{17}{20.4}{\rmdefault}{\mddefault}{\updefault}{\color[rgb]{0,0,0}Conflict Graph}%
}}}}
\put(11611,-3211){\makebox(0,0)[lb]{\smash{{\SetFigFont{17}{20.4}{\rmdefault}{\mddefault}{\updefault}{\color[rgb]{0,0,0}Final State}%
}}}}
\put(11611,-3886){\makebox(0,0)[lb]{\smash{{\SetFigFont{17}{20.4}{\rmdefault}{\mddefault}{\updefault}{\color[rgb]{0,0,0}Block}%
}}}}
\put(11611,-3616){\makebox(0,0)[lb]{\smash{{\SetFigFont{17}{20.4}{\rmdefault}{\mddefault}{\updefault}{\color[rgb]{0,0,0}Hash of Previous }%
}}}}
\end{picture}%

%% file: Background.tex
\section{System Model and Background}
\label{sec:model}
This section includes the commencement of notions related to this paper such as \bc{}, \scontract{s}, STMs and its execution model. Here, we limit our discussion to a well-known smart contracts platform, Ethereum. We improve the throughput by ensuring the concurrent execution of \scontract{s} using an efficient framework, optimistic STMs. 

\subsection{Blockchain and Smart Contracts}
Blockchain is a distributed and highly secure technology which stores the records into the block. It consists of multiple peers (or nodes), and each peer maintains decentralize distributed ledger that makes it publicly readable but tamper-proof. Peer executes some functions in the form of transactions. A transaction is a set of instructions executing in the memory. Bitcoin is a blockchain system which only maintains the balances while transferring the money from one account to another account in the distributed manner. Whereas, the popular blockchain system such as Ethereum maintains the state information as well. Here, transactions execute the atomic code known as a function of \scontract{}. Smart contract consists of one or more atomic-units or functions. In this paper, the atomic-unit contains multiple steps that have been executed by an efficient framework which is optimistic STMs.

\noindent
\textbf{Smart Contracts:} The transactions sent by clients to miners are part of a larger code called as \emph{\scontract{s}} that provide several complex services such as managing the system state, ensures rules, or credentials checking of the parties involved, etc. \cite{Dickerson+:ACSC:PODC:2017}. 
For better understanding of smart contract, we describe a simple auction contract from Solidity documentation \cite{Solidity}.\\
\textbf{Simple Auction Contract:} The functionality of simple auction contract is shown in \algoref{sa}. Where \Lineref{sa1} declares the contract, followed by public state variables as ``highestBidder, highestBid, and pendingReturn'' which records the state of the contract. A single owner of the contract initiates the auction by executing constructor ``SimpleAuction()'' method (omitted due to lack of space) in which function initialize bidding time as auctionEnd (\Lineref{sa3}). 
There can be any number of participants to bid. The bidders may get their money back whenever the highest bid is raised. For this, a public state variable declared at \Lineref{sa7} (pendingReturns) uses solidity built-in complex data type mapping to maps bidder addresses with unsigned integers (withdraw amount respective to bidder). Mapping can be seen as a hash table with key-value pair. This mapping uniquely identifies account addresses of the clients in the Ethereum blockchain. A bidder withdraws the amount of their earlier bid by calling withdraw() method \cite{Solidity}.

At \Lineref{sa8}, a contract function ``bid()'' is declared, which is called by bidders to bid in the auction. Next, ``auctionEnd'' variable is checked to identify whether the auction already called off or not. Further, bidders ``msg.value'' check to identify the highest bid value at \Lineref{sa11}. Smart contract methods can be aborted at any time via throw when the auction is called off, or bid value is smaller than current ``highestBid''. When execution reaches to \Lineref{sa14}, the ``bid()'' method recovers the current highest bidder data from mapping through the ``highestBidder'' address and updates the current bidder pending return amount. Finally, at \Lineref{sa16} and \Lineref{sa17}, it updates the new highest bidder and highest bid amount.

\begin{algorithm}
	\scriptsize
	\caption{SimpleAuction: It allows every bidder to send their bids throughout the bidding period.}	\label{alg:sa} 
	\begin{algorithmic}[1]
		\makeatletter\setcounter{ALG@line}{0}\makeatother
		\Procedure{Contract SimpleAuction}{} \label{lin:sa1}
		\State address public beneficiary;\label{lin:sa2}
		\State uint public auctionEnd;\label{lin:sa3}
		\State /*current state of the auction*/\label{lin:sa4}
		\State address public highestBidder;\label{lin:sa5}
		\State uint public highestBid;\label{lin:sa6}
		\State mapping(address $=>$ uint) pendingReturns; \label{lin:sa7} 
		\Function {}{}bid() public payable \label{lin:sa8}
		
		\If{(now $\geq$ auctionEnd)} 
		\State throw;\label{lin:sa10}
		\EndIf
		\If{(msg.value $<$ highestBid)} \label{lin:sa11}
		\State thorw;\label{lin:sa12}
		\EndIf
		\If{(highestBid != 0)}\label{lin:sa13}
		\State pendingReturns[highestBidder] += highestBid;\label{lin:sa14}
		\EndIf  \label{lin:sa15}
		\State highestBidder = msg.sender;\label{lin:sa16}
		\State highestBid = msg.value;\label{lin:sa17}
		\EndFunction
		\State // more operation definitions\label{lin:sa18}
		\EndProcedure
		
	\end{algorithmic}
\end{algorithm}


\noindent
\textbf{Software Transactional Memory Systems:} Following~\cite{tm-book,KuzSat:NI:TCS:2016}, we assume a system of $n$ processes/threads, $p_1,\ldots,p_n$ that access a collection of \emph{transactional objects} or \tobj{s} via atomic \emph{transactions}. Each transaction has a unique identifier. Within a transaction, processes can perform \emph{transactional operations or \mth{s}}: \textit{\begt{()}} that begins a transaction, \textit{\twrite}$(x,v)$ (or $w(x, v)$) that updates a \tobj{} $x$ with value $v$ in its local memory, \textit{\tread}$(x, v)$ (or $r(x, v)$) that tries to read  $x$ and returns value as $v$, \textit{\tryc}$()$ that tries to commit the transaction and returns $commit$ (or $\commit$) if it succeeds, and \textit{\trya}$()$ that aborts the transaction and returns $\abort$. 
Operations \textit{\tread{()}} and \textit{\tryc}$()$ may return $\abort{}$. 
Transaction $T_i$ starts with the first operation and completes when any of its operations return $\abort$ or $\commit$. 
For a transaction $T_k$, we denote all the \tobj{s} accessed by its read \op{s} and write operations as $\rs_k$ and $\ws_k$ respectively.  We denote all the \op{s} of a transaction $T_k$ as $\evts{T_k}$ or $evts_k$. 

\noindent
\textbf{History:}
A \emph{history} is a sequence of \emph{events}, i.e., a sequence of 
invocations and responses of transactional operations. The collection of events 
is denoted as $\evts{H}$. For simplicity, we only consider \emph{sequential} 
histories here: the invocation of each transactional operation is immediately followed by a matching response. Therefore, we treat each transactional operation as one atomic event and let $<_H$ denote the total order on the 
transactional operations incurred by $H$. 
We identify a 
history $H$ as tuple $\langle \evts{H},<_H \rangle$. 

We only consider \emph{well-formed} histories here, i.e., no transaction of a process begins before the previous transaction invocation has completed (either $commits$ or $aborts$). We also assume that every history has an initial \emph{committed} transaction $T_0$ that initializes all the t-objects with value $0$. The set of transactions that appear in $H$ is denoted by $\txns{H}$. The set of \emph{committed} (resp., \emph{aborted}) transactions in $H$ is denoted by $\comm{H}$ (resp., $\aborted{H}$). The set of \emph{incomplete} or \emph{live} transactions in $H$ is denoted by $\incomp{H} = \live{H} = (\txns{H}-\comm{H}-\aborted{H})$. 

We construct a \emph{complete history} of $H$, denoted as $\overline{H}$, by inserting $\trya_k(\abort)$ immediately after the last event of every transaction $T_k\in \live{H}$. But for $\tryc_i$ of transaction $T_i$, if it released the lock on first \tobj{} successfully that means updates made by $T_i$ is consistent so, $T_i$ will immediately return commit.
\cmnt{
\noindent
\textbf{Sub-history:} A \textit{sub-history} ($SH$) of a history ($H$) 
denoted as the tuple $\langle \evts{SH},$ $<_{SH}\rangle$ and is defined as: 
(1) $<_{SH} \subseteq <_{H}$; (2) $\evts{SH} \subseteq \evts{H}$; (3) If an 
event of a transaction $T_k\in\txns{H}$ is in $SH$ then all the events of $T_k$ 
in $H$ should also be in $SH$. 

For a history $H$, let $R$ be a subset of $\txns{H}$. Then $\shist{R}{H}$ denotes  the \ssch{} of $H$ that is formed  from the \op{s} in $R$.
}

\noindent
\textbf{\textit{Transaction Real-Time and Conflict order:}} For two transactions $T_k,T_m \in \txns{H}$, we say that  $T_k$ \emph{precedes} $T_m$ in the \emph{real-time order} of $H$, denoted as $T_k\prec_H^{RT} T_m$, if $T_k$ is complete in $H$ and the last event of $T_k$ precedes the first event of $T_m$ in $H$. If neither $T_k \prec_H^{RT} T_m$ nor $T_m \prec_H^{RT} T_k$, then $T_k$ and $T_m$ \emph{overlap} in $H$. We say that a history is \emph{\tseq} if all the transactions are ordered by real-time order. 

We say that $T_k, T_m$ are in conflict, denoted as $T_k\prec_H^{Conf} T_m$, if (1) $\tryc_k()<_H \tryc_m()$ and $wset(T_k) \cap wset(T_m) \neq\emptyset$; (2) $\tryc_k()<_H r_m(x,v)$, $x \in wset(T_k)$ and $v \neq \abort$; (3) $r_k(x,v)<_H \tryc_m()$, $x\in wset(T_m)$ and $v \neq \abort$. Thus, it can be seen that the conflict order is defined only on \op{s} that have successfully executed. We denote the corresponding \op{s} as conflicting. 

\noindent
\textbf{Valid and Legal histories:} A successful read $r_k(x, v)$ (i.e., $v \neq \abort$)  in a history $H$ is said to be \emph{\valid} if there exist a transaction $T_j$ that wrote $v$ to $x$ and \emph{committed} before $r_k(x,v)$. 
History $H$ is \valid{} if all its successful read \op{s} are \valid. 

We define $r_k(x, v)$'s \textit{\lastw{}} as the latest commit event $\commit_i$ preceding $r_k(x, v)$ in $H$ such that $x\in wset_i$ ($T_i$ can also be $T_0$). A successful read \op{} $r_k(x, v)$ (i.e., $v \neq \abort$), is said to be \emph{\legal{}}  if the transaction containing $r_k$'s \lastw{} also writes $v$ onto $x$.
 The history $H$ is \legal{} if all its successful read \op{s} are \legal.  From the definitions we get that if $H$ is \legal{} then it is also \valid.

%

\noindent
\textbf{Notions of Equivalence:} Two histories $H$ and $H'$ are \emph{equivalent} if they have the same set of events. We say two histories $H, H'$ are \emph{multi-version view equivalent} \cite[Chap. 5]{WeiVoss:TIS:2002:Morg} or \emph{\mvve} if (1) $H, H'$ are valid histories and (2) $H$ is equivalent to $H'$. 

Two histories $H, H'$ are \emph{view equivalent} \cite[Chap. 3]{WeiVoss:TIS:2002:Morg} or \emph{\vie} if (1) $H, H'$ are legal histories and (2) $H$ is equivalent to $H'$. By restricting to \legal{} histories, view equivalence does not use multi-versions. 

Two histories $H, H'$ are \emph{conflict equivalent} \cite[Chap. 3]{WeiVoss:TIS:2002:Morg} or \emph{\ce} if (1) $H, H'$ are legal histories and (2) conflict in $H, H'$ are the same, i.e., $conf(H) = conf(H')$. Conflict equivalence like view equivalence does not use multi-versions and restricts itself to \legal{} histories. 

\noindent
\textbf{VSR, MVSR, and CSR:} A history $H$ is said to VSR (or View Serializable) \cite[Chap. 3]{WeiVoss:TIS:2002:Morg}, if there exist a serial history $S$ such that $S$ is view equivalent to $H$. But it maintains only one version corresponding to each \tobj{}.

So, MVSR (or Multi-Version View Serializable) came into picture which maintains multiple version corresponding to each \tobj. A history $H$ is said to MVSR \cite[Chap. 5]{WeiVoss:TIS:2002:Morg}, if there exist a serial history $S$ such that $S$ is multi-version view equivalent to $H$. It can be proved that verifying the membership of VSR as well as MVSR in databases is NP-Complete \cite{Papad:1979:JACM}. To circumvent this issue, researchers in databases have identified an efficient sub-class of VSR, called CSR based
on the notion of conflicts. The membership of CSR can be verified in polynomial time using conflict graph characterization. 

A history $H$ is said to CSR (or Conflict Serializable) \cite[Chap. 3]{WeiVoss:TIS:2002:Morg}, if there exist a serial history $S$ such that $S$ is conflict equivalent to $H$.
\\
\noindent
\textbf{Serializability and Opacity:} 
Serializability\cite{Papad:1979:JACM} is a commonly used criterion in databases. But it is not suitable for STMs as it does not consider the correctness of \emph{aborted} transactions as shown by Guerraoui \& Kapalka \cite{GuerKap:Opacity:PPoPP:2008}. Opacity, on the other hand, considers the correctness of \emph{aborted} transactions as well.

A history $H$ is said to be \textit{opaque} \cite{GuerKap:Opacity:PPoPP:2008,tm-book} if it is \valid{} and there exists a t-sequential legal history $S$ such that (1) $S$ is equivalent to complete history $\overline{H}$ and (2) $S$ respects $\prec_{H}^{RT}$, i.e., $\prec_{H}^{RT} \subset \prec_{S}^{RT}$.  By requiring $S$ being equivalent to $\overline{H}$, opacity treats all the incomplete transactions as aborted. 

\cmnt{
Along same lines, a \valid{} history $H$ is said to be \textit{strictly serializable} if $\shist{\comm{H}}{H}$ is opaque. Unlike opacity, strict serializability does not include aborted or incomplete transactions in the global serialization order. An opaque history $H$ is also strictly serializable: a serialization of $\shist{\comm{H}}{H}$ is simply the subsequence of a serialization of $H$ that only contains transactions in $\comm{H}$. 
}

\cmnt{
\noindent 
\textbf{VSR, MVSR, and CSR:} VSR (or View Serializability) is a correctness-criterion same as opacity but it does not consider aborted transactions. When the protocol (such as MVTO) maintains multiple versions corresponding to each \tobj{} then a commonly used correctness criterion in databases is MVSR (or Multi-Version View Serializability). It can be proved that verifying the membership of VSR as well as MVSR in databases is NP-Complete \cite{Papad:1979:JACM}. To circumvent this issue, researchers in databases have identified an efficient sub-class of VSR, called CSR (or conflict-serializability), based
on the notion of conflicts. The membership of CSR can be verified in polynomial time using conflict graph characterization. 
}

\noindent
\textbf{Linearizability:} A history $H$ is linearizable \cite{HerlihyandWing:1990:LCC:ACM} if (1) The invocation and response events can be reordered to get a valid sequential history.
(2) The generated sequential history satisfies the object’s sequential specification. (3) If a response event precedes an invocation event in the original history, then this should be preserved in the sequential reordering.

\noindent
\textbf{Lock Freedom:} An algorithm is said to be lock-free\cite{HerlihyShavit:Progress:Opodis:2011} if the program threads are run for a sufficiently long time, at least one of the threads makes progress. It allows individual threads to starve but guarantees system-wide throughput.

\ignore{


\begin{enumerate}
	\item BTO 
	\begin{itemize}
		\item Conflicts.
	\end{itemize}
	\item MVTO 
	\begin{itemize}
		\item Conflicts.
	\end{itemize}
\end{enumerate}
}

%% file: objective.tex
\section{Requirements of Concurrent Miner, Validator and Block Graph}
\label{sec:req-miner_val}

This section describes the requirements of concurrent miner, validator and block graph to ensure correct concurrent execution of the \scontract transactions. 


\subsection{Requirements of the Concurrent Miner}
The miner process invokes several threads to concurrently execute the \scontract transactions or \au{s}. With the proposed optimistic execution approach, each miner thread invokes an \au as a transaction. 

The miner should ensure the correct concurrent execution of the smart contract transactions. The incorrect concurrent execution (or consistency issues) may occur when concurrency involved. Any inconsistent read may leads system to divide by zero, infinite loops, crash failure etc. All smart contract transactions take place within a virtual machine \cite{Dickerson+:ACSC:PODC:2017}. 
When miner executes the smart contract transactions concurrently on the virtual machine then infinite loop and inconsistent read may occur. So, to ensure the correct concurrent execution, the miner should satisfy the correctness-criterion as opacity \cite{GuerKap:Opacity:PPoPP:2008}.

To achieve better efficiency, sometimes we need to adapt the non-virtual machine environment which necessitates with the safeguard of transactions. There as well miner needs to satisfies the correctness-criterion as opacity to ensure the correct concurrent execution of smart contract transactions. 

\begin{requirement}
	Any history $H_m$ generated by concurrent miner should satisfy opacity.
\end{requirement}
\cmnt{
\begin{proof}
	Here, miner executes the smart contract concurrently with the help of optimistic STM protocols (BTO and MVTO). Internally, BTO and MVTO \cite{Kumar+:MVTO:ICDCN:2014} protocol ensures opacity. So, history $H_m$ generated Concurrent miner satisfies opacity.
\end{proof}
}



Concurrent miner maintains a BG and provides it to concurrent validators which ensures the dependency order among the conflicting transactions. As we discussed in \secref{intro}, if concurrent miner will not maintain the BG then a valid block may get rejected by the concurrent validator. 

\subsection{Requirements of the Concurrent Validator}
The correct concurrent execution by validator should be equivalent to some serial execution. The serial order can be obtained by applying the topological sort on the BG provided by the concurrent miner. BG gives partial order among the transactions while restricting the dependency order same as the concurrent miner. So validator executes those transactions concurrently which are not having any dependency among them with the help of BG. Validator need not have to worry about any concurrency control issues because BG ensures conflicting transactions never execute concurrently.

\subsection{Requirements of the Block Graph}

As explained above, the miner generates a BG to capture the dependencies between the smart contract transactions which is used by the validator to concurrently execute the transactions again later. The validator executes those transactions concurrently which do not have any path (implying dependency) between them. Thus the execution by the validator is  given by a topological sort on the BG. 

Now it is imperative that the execution history generated by the validator, $H_v$ is `equivalent' to the history generated by the miner, $H_m$. The precise equivalence depends on the STM protocol followed by the miners and validators. If the miner uses Multi-version STM such as MVTO then the equivalence between $H_v$ and $H_m$ is \mvve. In this case, the graph generated by the miner would be multi-version serialization graph \cite[Chap. 5]{WeiVoss:TIS:2002:Morg}. 

On the other hand, if the miner uses single version STM such as BTO then the equivalence between $H_v$ and $H_m$ is view-equivalence (\vie) which can be approximated by conflict-equivalence (\ce). Hence, in this case, the graph generated by the miner would be conflict graph \cite[Chap. 3]{WeiVoss:TIS:2002:Morg}.

%% file: pm.tex
\section{Proposed Mechanism}
\label{sec:pm}
This section presents the methods of lock-free concurrent block graph library followed by concurrent execution of smart contract transactions by miner and validator.
\subsection{Lock-free Concurrent Block Graph}
\label{subsec:bg}
\noindent
\textbf{Data Structure of Lock-free Concurrent Block Graph:} We use \emph{adjacency list} to maintain the block graph BG(V, E) as shown in \figref{confg} (a). Where V is set of vertices (or \vnode{s}) which are stored in the vertex list (or \vl{}) in increasing order of timestamp between two sentinel node \vh{} (-$\infty$) and \vt{} (+$\infty$). Each vertex node (or \vnode) contains $\langle ts = i, AU_{id} = id, \inc{} = 0, \vn{} = nil, \en{} = nil\rangle$. Where $i$ is a unique timestamp (or $ts$) of committed transactions $T_i$. $AU_{id}$ is the $id$ of \au{} which is executed by transaction $T_i$. To maintain the indegree count of each \vnode{} we initialize \inc{} as 0. \vn{} and \en{} initializes as $nil$. 

Here, E is a set of edges which maintains all the conflicts of \vnode{} in the edge list (or \el) as shown in \figref{confg} (a). \el{} stores \enode{s} (or conflicting transaction nodes say $T_j$) in increasing order of timestamp (or $ts$) between two sentinel nodes \eh{} (-$\infty$) and \et{} (+$\infty$). 
\vspace{-.3cm}
\begin{figure}[H]
	\centering
		\scalebox{.31}{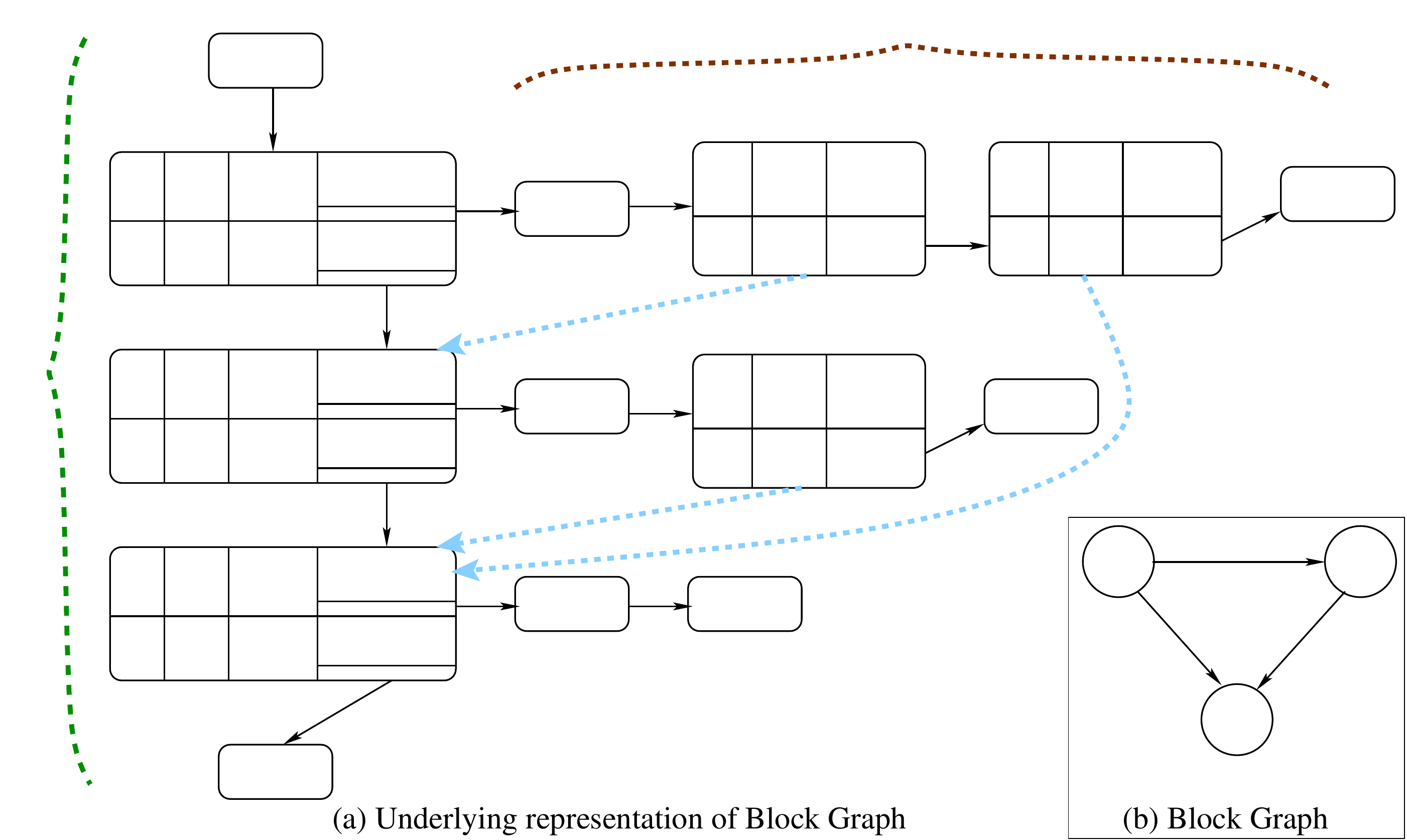}
		\centering
		\caption{Pictorial representation of Block Graph}
		\label{fig:confg}
\end{figure}
\vspace{-.4cm}

\cmnt{
\begin{figure*}
	\centering
	\begin{minipage}[b]{0.3\textwidth}
		\scalebox{.31}{\input{figs/graphs.pdf_t}}
		\centering
		\caption{Pictorial representation of Block Graph}
		\label{fig:confg}
	\end{minipage}
	\hfill
	\begin{minipage}[b]{0.5\textwidth}
		\scalebox{.32}{\input{figs/cminer.pdf_t}}
		\centering
		\caption{Execution of Concurrent Miner}
		\label{fig:cminer}
	\end{minipage}   
\end{figure*}
}
Edge node (or \enode{}) contains $\langle$\emph{ts = j, vref}, \en{} = $nil$ $\rangle$. Here, $j$ is a unique timestamp (or $ts$) of \emph{committed} transaction $T_j$ which is having conflict with $T_i$ and $ts(T_i)$ is less than $ts(T_j)$. To maintain the acyclicity of the BG, we add a conflict edge from lower timestamp transaction to higher timestamp transaction i.e. conflict edge is from $T_i$ to $T_j$ in the BG. \figref{confg} (b) illustrates this using three transactions with timestamp 0, 5, and 10, which maintain the acyclicity while adding an edge from lower to higher timestamp. \emph{Vertex node reference (or vref)} keeps the reference of its own vertex which is present in the \vl. \en{} initializes as $nil$. 

Block graph generated by the concurrent miner which helps to execute the validator concurrently and deterministically through lock-free graph library methods. Lock-free graph library consists of five methods as follows: \emph{addVert(), addEdge(), searchLocal(), searchGlobal()} and \emph{decInCount()}. 

\noindent
\textbf{Lock-free Graph Library Methods Accessed by Concurrent Miner:} Concurrent miner uses addVert() and addEdge() methods of lock-free graph library to build a BG. When concurrent miner wants to add a node in the BG then first it calls addVert() method. addVert() method identifies the correct location of that node (or \vgn{}) in the \vl{}. If \vgn{} is not part of \vl{} then it creates the node and adds it into \vl{} in lock-free manner with the help of atomic compare and swap operation. 

After successful addition of \vnode{} in the BG concurrent miner calls addEdge() method to add the conflicting node (or \egn{}) corresponding to \vnode{} in the \el{}. First, addEdge() method identifies the correct location of \egn{} in the \el{} of corresponding \vnode{}. If \egn{} is not part of \el{} then it creates the node and adds it into \el{} of \vnode{} in lock-free manner with the help of atomic compare and swap operation. After successful addition of \enode{} in the \el{} of \vnode{}, it increment the \inc{} of \enode.$vref$ (to maintain the indegree count) node which is present in the \vl{}.

\noindent
\textbf{Lock-free Graph Library Methods Accessed by Concurrent Validator:} Concurrent validator uses searchLocal(), searchGlobal() and decInCount() methods of lock-free graph library. First, concurrent validator thread calls searchLocal() method to identify the source node (having indegree (or \inc) 0) in its local \cachel{} (or thread local memory). If any source node exist in the local \cachel{} with \inc{} 0 then it sets \inc{} field to be -1 atomically to claim the ownership of the node.

If the source node does not exists in the local \cachel{} then concurrent validator thread calls searchGlobal() method to identify the source node in the BG. If any source node exists in the BG then it will do the same process as done by searchLocal(). After that validator thread calls the decInCount() to decreases the \inc{} of all the conflicting nodes atomically which are present in the \el{} of corresponding source node. While decrementing the \inc{} of each conflicting nodes in the BG, it again checks if any conflicting node became a source node then it adds that node into its local \cachel{} to optimize the search time of identifying the next source node. Due to lack of space, please refer \apnref{apnbg} to get the complete details with the algorithm of lock-free graph library methods.


\subsection{Concurrent Miner}
\label{subsec:cminer}
Smart contracts in \bc{} are executed in two different context. First, by \Miner{} to propose a new block and after that by multiple \Validator{s} to verify the block proposed by \Miner. In this subsection, we describe how miner executes the \SContract{s} concurrently. \tikz \node[circle,scale=.5,color=black, fill=pink]{\textbf{1}}; \emph{Concurrent miner} gets the set of transactions from the \emph{distributed shared memory} as shown in \figref{cminer}. Each transaction associated with the functions (or \au{s}) of smart contracts. To run the \SContract{s} concurrently we have faced the challenge to identify the conflicting transactions at run-time because \SContract{} language are Turing-complete. Two transactions are in conflict if they are accessing common shared data-objects and at least one of them perform write operation on it. \tikz \node[circle,scale=.5,color=black, fill=pink]{\textbf{2}}; In \conminer{}, conflicts are identified at run-time with the help of efficient framework provided by optimistic software transactional memory system (STMs). STMs access the shared data-objects called as \tobj{s}. Each shared \tobj{} having initial state (or IS) which modified by the \au{s} and change IS to some other valid state. Eventually, it reaches to final state (or FS) at the end of block creation. As shown in \algoref{cminer}, first, each transaction $T_i$ gets the unique timestamp $i$ from STM.begin() at \Lineref{cminer5}. Then transaction $T_i$ executes the \au{} of \SContract{s}. \emph{Atomic-unit} consists of multiple steps such as $read$ and $write$ on shared \tobj{s} as $x$. Internally, these $read$ and $write$ steps are handled by the STM.read() and STM.write(), respectively. At \Lineref{cminer9}, if current \au{} step (or curStep) is $read(x)$ then it calls the STM.read(x). Internally, STM.read() identify the shared \tobj{} $x$ from transactional memory (or TM) and validate it. If validation is successful then it gets the value as $v$ at \Lineref{cminer10} and execute the next step of \au{} otherwise re-execute the \au{} if $v$ is $abort$ at \Lineref{cminer11}. 
\vspace{-.3cm}
\begin{figure}[H]
	\scalebox{.32}{\input{figs/cminer.pdf_t}}
	\centering
	\caption{Execution of Concurrent Miner}
	\label{fig:cminer}
\end{figure}
\vspace{-.4cm}

\cmnt{
\begin{figure*}
	\centering
	\centerline{\scalebox{0.45}{\input{figs/cminer.pdf_t}}}
	\caption{Execution of Concurrent Miner}
	\label{fig:cminer}
\end{figure*}
}
If curStep is $write(x)$ at \Lineref{cminer14} then it calls the STM.write(x). Internally, STM.write() stores the information corresponding to the shared \tobj{} $x$ into transaction local log (or \txlog) in write-set (or $wset_i$) for transaction $T_i$. We use an optimistic approach in which effect of the transaction will reflect onto the TM after the successful STM.tryC(). If validation is successful for all the $wset_i$ of transaction $T_i$ in STM.tryC() i.e. all the changes made by the $T_i$ is consistent then it updates the TM otherwise re-execute the \au{} if $v$ is $abort$ at \Lineref{cminer23}. After successful validation of STM.tryC(), it also maintains the conflicting transaction of $T_i$ into conflict list in TM.
\cmnt{
\begin{figure*}
	\centerline{\scalebox{0.45}{\input{figs/graphs.pdf_t}}}
	\caption{Pictorial representation of \confg{}, CG}
	\label{fig:confg}
\end{figure*}
}
\cmnt{
Once the transaction commits it stores its conflicts in the form of \CG{}. In order to store the \emph{Conflict Graph} or \confg \emph{(CG(V, E))}, we are maintaining \emph{adjacency list}. In which all the vertices (or \vnode{s}) are stored in the vertex list (or \vl{}) in increasing order of timestamp between the two sentinal node \vh{} (-$\infty$) and \vt{} (+$\infty$). The structure of the vertex node (or \vnode) is as $\langle ts_i, AU_{id}, \inc{}, \vn, \en\rangle$. Where $ts$ is a unique timestamp $i$ of committed transactions $T_i$ assign at the beginning of it. $AU_{id}$ is the $id$ of \au{} which is executed by transaction $T_i$. To maintain the indegree count of each \vnode{} we use \inc{} initialize as 0. \vn{} points to the next vertex of the \vl. For corresponding to each \vnode{}, it maintains all the conflicts in the edge list (\el) as shown in \figref{confg}. \el{} stores \enode{} (or conflicting transaction nodes) belonging to \emph{vNode} stores in increasing order of timestamp (or $ts$) in between the two sentinal nodes \eh{} (-$\infty$) and \et{} (+$\infty$). The structure of \enode{} is as $\langle$\emph{$ts_j$, vref}, \en{} $\rangle$. $ts_j$ is a unique timestamp $j$ of \emph{committed} transaction $T_j$ which had a conflict with $T_i$. \emph{Vertex node reference (or vref)} keeps the reference of its own vertex which is present in the \vl. \en{} points the next \enode{} of the \el{}.
}
\begin{algorithm}
	\scriptsize 
	\caption{\cminer{(\aul, STM)}: Concurrently $m$ threads are executing atomic-units of smart contract from \aul{}(or list of atomic-units) with the help of STM.}
	\label{alg:cminer}	
	\begin{algorithmic}[1]
		\makeatletter\setcounter{ALG@line}{0}\makeatother
		\Procedure{\cminer{(\aul, STM)}}{}\label{lin:cminer1}
		\State curAU $\gets$ $curInd$.$get\&Inc(\aul)$; \label{lin:cminer2}
		\State /*curAU is the current atomic-unit taken from the \aul*/
		\State /*Execute until all the atomic-units successfully completed*/\label{lin:cminer3}
		\While{(curAU $<$ size\_of(\aul))}\label{lin:cminer4}
		\State $T_i$ $\gets$ STM.\begtrans{()};/*Create a new transaction $T_i$ with timestamp $i$*/\label{lin:cminer5}
		\While{(curAU.steps.hasNext())} /*Assume that curAU is a list of steps*/\label{lin:cminer6}
		\State curStep = currAU.steps.next(); /*Get the next step to execute*/\label{lin:cminer7}
		\Switch{(curStep)}\label{lin:cminer8}
		\EndSwitch
		\Case{read($x$):}\label{lin:cminer9}
		\State $v$ $\gets$ STM.\readi{($x$)}; /*Read \tobj{} $x$ from a shared memory*/\label{lin:cminer10}
		\If{($v$ == $abort$)}\label{lin:cminer11}
		\State goto \Lineref{cminer5};\label{lin:cminer12}
		\EndIf\label{lin:cminer13}
		\EndCase
		\Case{write($x, v$):} \label{lin:cminer14}
		\State /*Write \tobj{} $x$ into $T_i$ local memory with value $v$*/\label{lin:cminer15}
		\State STM.$write_i$($x, v$); \label{lin:cminer16}
		\EndCase
		\Case{default:}\label{lin:cminer17}
		\State /*Neither read from or write to a shared memory \tobj{}*/\label{lin:cminer18}
		\State execute curStep;\label{lin:cminer19}
		\EndCase
		\EndWhile\label{lin:cminer20}
		\State /*Try to commit the current transaction $T_i$ and update the \cl{[i]}*/\label{lin:cminer21}
		\State $v$ $\gets$ \tryc{$_i$()}; \label{lin:cminer22}
		\If{($v == abort$)}\label{lin:cminer23}
		\State goto \Lineref{cminer5};\label{lin:cminer24}
		\EndIf			\label{lin:cminer25}
		\State Create \vnode{} with $\langle$\emph{$i$, $AU_{id}$, 0, nil, nil}$\rangle$ as a vertex of Block Graph;	\label{lin:cminer26}
		\State BG(\emph{vNode}, STM);		\label{lin:cminer27}
		\State curAU $\gets$ $curInd$.$get\&Inc(\aul)$; \label{lin:cminer28}
		\EndWhile	\label{lin:cminer29}
		\EndProcedure\label{lin:cminer30}
		
	\end{algorithmic}
\end{algorithm}

\tikz \node[circle,scale=.5,color=black, fill=pink]{\textbf{3}}; Once the transaction commits, it stores the conflicts in the block graph (or BG). To maintain the BG it calls addVert() and addEdge() methods of the lock-free graph library. The internal details of addVert() and addEdge() methods are explained in \subsecref{bg}. 
\tikz \node[circle,scale=.5,color=black, fill=pink]{\textbf{4}}; Once the transactions successfully executed the \au{s} and completed with the construction of BG then \conminer{} compute the hash of the previous block. Eventually, \tikz \node[circle,scale=.5,color=black, fill=pink]{\textbf{5}}; \conminer{} propose a block which consists of set of transactions, BG, final state of each shared \tobj{s}, hash of the previous block of the blockchain and \tikz \node[circle,scale=.5,color=black, fill=pink]{\textbf{6}}; send it to all other existing node in the \emph{distributed shared memory} to validate it as shown in \figref{cminer}. 


\subsection{Concurrent Validator}
\label{subsec:cvalidator}
Concurrent validator validates the block proposed by the concurrent miner. 
It executes the set of transactions concurrently and deterministically with the help of block graph given by the \conminer{}. BG consists of dependency among the conflicting transactions that restrict them to execute serially whereas non-conflicting transactions can run concurrently.
In \convalidator{} multiple threads are executing the \au{s} of \SContract{s} concurrently by \emph{executeCode()} method at \Lineref{sl4} and \Lineref{sl41} with the help of searchLocal(), and searchGlobal() and decInCount() methods of lock-free graph library at \Lineref{val5}, \Lineref{val16} and (\Lineref{val8}, \Lineref{val20}) respectively. 
The functionality of these lock-free graph library methods are explained in \subsecref{bg}. 



After the successful execution of all the \au{s}, \convalidator{} compares its computed final state of each shared data-objects with the final states given by the \conminer{}. If the final state matches for all the shared data-objects then the block proposed by the \conminer{} is valid. Finally, the block is appended to the blockchain and respective \conminer{} is rewarded.
\begin{algorithm}
\scriptsize
	\caption{\cvalidator{(\aul, BG)}: Concurrently $V$ threads are executing atomic-units of smart contract with the help of BG given by the miner.}
	\begin{algorithmic}[1]
		\makeatletter\setcounter{ALG@line}{31}\makeatother
		\Procedure{\cvalidator{(\aul, BG)}}{}		
		\State /*Execute until all the atomic-units successfully completed*/ \label{lin:val1}
		\While{(\nc{} $<$ size\_of(\aul))} \label{lin:val2}
		\While{(\cachel{}.hasNext())} /*First search into thread local \cachel*/\label{lin:val3}
		\State cacheVer $\gets$ \cachel{}.next(); \label{lin:val4}
		\State cacheVertex $\gets$ \searchl{(cacheVer, $AU_{id}$)};\label{lin:val5}
		\State  \exec{($AU_{id}$)};\label{lin:sl4} /*Execute the atomic-unit of cacheVertex*/
		\While{(cacheVertex)}\label{lin:val7}
		\State cacheVertex $\gets$ decInCount(cacheVertex);\label{lin:val8}
	
		\EndWhile\label{lin:val10}
		\State Remove the current node (or cacheVertex) from local \cachel; \label{lin:val12}
		\EndWhile\label{lin:val13}
		\State vexNode $\gets$ \searchg{(BG, $AU_{id}$)}; /*Search into the BG*/\label{lin:val16}
		\State  \exec{($AU_{id}$)};\label{lin:sl41} /*Execute the atomic-unit of vexNode*/		
		\While{(verNode)}\label{lin:val19}
		\State verNode $\gets$ decInCount(verNode);\label{lin:val20}
		\EndWhile\label{lin:val22}
		\EndWhile\label{lin:val27}
		\EndProcedure
	\end{algorithmic}
\end{algorithm}
\begin{theorem}
	All the dependencies between the conflicting nodes are captured in the BG.
\end{theorem}

%% file: figs/graphs.pdf_t
\begin{picture}(0,0)%
\includegraphics{figs/graphs.pdf}%
\end{picture}%
\setlength{\unitlength}{4144sp}%
\begingroup\makeatletter\ifx\SetFigFont\undefined%
\gdef\SetFigFont#1#2#3#4#5{%
  \reset@font\fontsize{#1}{#2pt}%
  \fontfamily{#3}\fontseries{#4}\fontshape{#5}%
  \selectfont}%
\fi\endgroup%
\begin{picture}(12804,7641)(1654,-8758)
\put(3781,-1726){\makebox(0,0)[lb]{\smash{{\SetFigFont{17}{20.4}{\rmdefault}{\mddefault}{\updefault}{\color[rgb]{0,0,0}$-\infty$}%
}}}}
\put(6571,-3076){\makebox(0,0)[lb]{\smash{{\SetFigFont{17}{20.4}{\rmdefault}{\mddefault}{\updefault}{\color[rgb]{0,0,0}$-\infty$}%
}}}}
\put(6571,-4876){\makebox(0,0)[lb]{\smash{{\SetFigFont{17}{20.4}{\rmdefault}{\mddefault}{\updefault}{\color[rgb]{0,0,0}$-\infty$}%
}}}}
\put(6571,-6676){\makebox(0,0)[lb]{\smash{{\SetFigFont{17}{20.4}{\rmdefault}{\mddefault}{\updefault}{\color[rgb]{0,0,0}$-\infty$}%
}}}}
\put(8146,-3436){\makebox(0,0)[lb]{\smash{{\SetFigFont{17}{20.4}{\rmdefault}{\mddefault}{\updefault}{\color[rgb]{0,0,0}5}%
}}}}
\put(8056,-2761){\makebox(0,0)[lb]{\smash{{\SetFigFont{17}{20.4}{\rmdefault}{\mddefault}{\updefault}{\color[rgb]{0,0,1}$ts$}%
}}}}
\put(8551,-2761){\makebox(0,0)[lb]{\smash{{\SetFigFont{17}{20.4}{\rmdefault}{\mddefault}{\updefault}{\color[rgb]{0,0,1}$vref$}%
}}}}
\put(9316,-2761){\makebox(0,0)[lb]{\smash{{\SetFigFont{17}{20.4}{\rmdefault}{\mddefault}{\updefault}{\color[rgb]{0,0,1}$eNext$}%
}}}}
\put(3961,-8251){\makebox(0,0)[lb]{\smash{{\SetFigFont{17}{20.4}{\rmdefault}{\mddefault}{\updefault}{\color[rgb]{0,0,0}$+\infty$}%
}}}}
\put(8236,-6721){\makebox(0,0)[lb]{\smash{{\SetFigFont{17}{20.4}{\rmdefault}{\mddefault}{\updefault}{\color[rgb]{0,0,0}$+\infty$}%
}}}}
\put(10936,-4921){\makebox(0,0)[lb]{\smash{{\SetFigFont{17}{20.4}{\rmdefault}{\mddefault}{\updefault}{\color[rgb]{0,0,0}$+\infty$}%
}}}}
\put(8146,-5371){\makebox(0,0)[lb]{\smash{{\SetFigFont{17}{20.4}{\rmdefault}{\mddefault}{\updefault}{\color[rgb]{0,0,0}10}%
}}}}
\put(8056,-4696){\makebox(0,0)[lb]{\smash{{\SetFigFont{17}{20.4}{\rmdefault}{\mddefault}{\updefault}{\color[rgb]{0,0,1}$ts$}%
}}}}
\put(8551,-4696){\makebox(0,0)[lb]{\smash{{\SetFigFont{17}{20.4}{\rmdefault}{\mddefault}{\updefault}{\color[rgb]{0,0,1}$vref$}%
}}}}
\put(9316,-4696){\makebox(0,0)[lb]{\smash{{\SetFigFont{17}{20.4}{\rmdefault}{\mddefault}{\updefault}{\color[rgb]{0,0,1}$eNext$}%
}}}}
\put(13636,-2986){\makebox(0,0)[lb]{\smash{{\SetFigFont{17}{20.4}{\rmdefault}{\mddefault}{\updefault}{\color[rgb]{0,0,0}$+\infty$}%
}}}}
\put(10846,-3436){\makebox(0,0)[lb]{\smash{{\SetFigFont{17}{20.4}{\rmdefault}{\mddefault}{\updefault}{\color[rgb]{0,0,0}10}%
}}}}
\put(10756,-2761){\makebox(0,0)[lb]{\smash{{\SetFigFont{17}{20.4}{\rmdefault}{\mddefault}{\updefault}{\color[rgb]{0,0,1}$ts$}%
}}}}
\put(11251,-2761){\makebox(0,0)[lb]{\smash{{\SetFigFont{17}{20.4}{\rmdefault}{\mddefault}{\updefault}{\color[rgb]{0,0,1}$vref$}%
}}}}
\put(12016,-2761){\makebox(0,0)[lb]{\smash{{\SetFigFont{17}{20.4}{\rmdefault}{\mddefault}{\updefault}{\color[rgb]{0,0,1}$eNext$}%
}}}}
\put(11701,-6316){\makebox(0,0)[lb]{\smash{{\SetFigFont{17}{20.4}{\rmdefault}{\mddefault}{\updefault}{\color[rgb]{0,0,0}$T_0$}%
}}}}
\put(13906,-6316){\makebox(0,0)[lb]{\smash{{\SetFigFont{17}{20.4}{\rmdefault}{\mddefault}{\updefault}{\color[rgb]{0,0,0}$T_5$}%
}}}}
\put(12781,-7756){\makebox(0,0)[lb]{\smash{{\SetFigFont{17}{20.4}{\rmdefault}{\mddefault}{\updefault}{\color[rgb]{0,0,0}$T_{10}$}%
}}}}
\put(3241,-2896){\makebox(0,0)[lb]{\smash{{\SetFigFont{17}{20.4}{\rmdefault}{\mddefault}{\updefault}{\color[rgb]{0,0,1}$AU$}%
}}}}
\put(4681,-2851){\makebox(0,0)[lb]{\smash{{\SetFigFont{17}{20.4}{\rmdefault}{\mddefault}{\updefault}{\color[rgb]{0,0,1}$eNext$}%
}}}}
\put(4681,-3436){\makebox(0,0)[lb]{\smash{{\SetFigFont{17}{20.4}{\rmdefault}{\mddefault}{\updefault}{\color[rgb]{0,0,1}$vNext$}%
}}}}
\put(2836,-3526){\makebox(0,0)[lb]{\smash{{\SetFigFont{17}{20.4}{\rmdefault}{\mddefault}{\updefault}{\color[rgb]{0,0,0}0}%
}}}}
\put(4096,-3526){\makebox(0,0)[lb]{\smash{{\SetFigFont{17}{20.4}{\rmdefault}{\mddefault}{\updefault}{\color[rgb]{0,0,0}0}%
}}}}
\put(2746,-2896){\makebox(0,0)[lb]{\smash{{\SetFigFont{17}{20.4}{\rmdefault}{\mddefault}{\updefault}{\color[rgb]{0,0,1}$ts$}%
}}}}
\put(3781,-2896){\makebox(0,0)[lb]{\smash{{\SetFigFont{17}{20.4}{\rmdefault}{\mddefault}{\updefault}{\color[rgb]{0,0,1}$inCnt$}%
}}}}
\put(3376,-3526){\makebox(0,0)[lb]{\smash{{\SetFigFont{17}{20.4}{\rmdefault}{\mddefault}{\updefault}{\color[rgb]{0,0,0}1}%
}}}}
\put(3241,-4696){\makebox(0,0)[lb]{\smash{{\SetFigFont{17}{20.4}{\rmdefault}{\mddefault}{\updefault}{\color[rgb]{0,0,1}$AU$}%
}}}}
\put(4681,-4651){\makebox(0,0)[lb]{\smash{{\SetFigFont{17}{20.4}{\rmdefault}{\mddefault}{\updefault}{\color[rgb]{0,0,1}$eNext$}%
}}}}
\put(4681,-5236){\makebox(0,0)[lb]{\smash{{\SetFigFont{17}{20.4}{\rmdefault}{\mddefault}{\updefault}{\color[rgb]{0,0,1}$vNext$}%
}}}}
\put(2836,-5326){\makebox(0,0)[lb]{\smash{{\SetFigFont{17}{20.4}{\rmdefault}{\mddefault}{\updefault}{\color[rgb]{0,0,0}5}%
}}}}
\put(3376,-5326){\makebox(0,0)[lb]{\smash{{\SetFigFont{17}{20.4}{\rmdefault}{\mddefault}{\updefault}{\color[rgb]{0,0,0}2}%
}}}}
\put(2746,-4696){\makebox(0,0)[lb]{\smash{{\SetFigFont{17}{20.4}{\rmdefault}{\mddefault}{\updefault}{\color[rgb]{0,0,1}$ts$}%
}}}}
\put(3781,-4696){\makebox(0,0)[lb]{\smash{{\SetFigFont{17}{20.4}{\rmdefault}{\mddefault}{\updefault}{\color[rgb]{0,0,1}$inCnt$}%
}}}}
\put(3241,-6496){\makebox(0,0)[lb]{\smash{{\SetFigFont{17}{20.4}{\rmdefault}{\mddefault}{\updefault}{\color[rgb]{0,0,1}$AU$}%
}}}}
\put(4681,-6451){\makebox(0,0)[lb]{\smash{{\SetFigFont{17}{20.4}{\rmdefault}{\mddefault}{\updefault}{\color[rgb]{0,0,1}$eNext$}%
}}}}
\put(4681,-7036){\makebox(0,0)[lb]{\smash{{\SetFigFont{17}{20.4}{\rmdefault}{\mddefault}{\updefault}{\color[rgb]{0,0,1}$vNext$}%
}}}}
\put(2836,-7126){\makebox(0,0)[lb]{\smash{{\SetFigFont{17}{20.4}{\rmdefault}{\mddefault}{\updefault}{\color[rgb]{0,0,0}10}%
}}}}
\put(4096,-7126){\makebox(0,0)[lb]{\smash{{\SetFigFont{17}{20.4}{\rmdefault}{\mddefault}{\updefault}{\color[rgb]{0,0,0}2}%
}}}}
\put(2746,-6496){\makebox(0,0)[lb]{\smash{{\SetFigFont{17}{20.4}{\rmdefault}{\mddefault}{\updefault}{\color[rgb]{0,0,1}$ts$}%
}}}}
\put(3781,-6496){\makebox(0,0)[lb]{\smash{{\SetFigFont{17}{20.4}{\rmdefault}{\mddefault}{\updefault}{\color[rgb]{0,0,1}$inCnt$}%
}}}}
\put(4096,-5326){\makebox(0,0)[lb]{\smash{{\SetFigFont{17}{20.4}{\rmdefault}{\mddefault}{\updefault}{\color[rgb]{0,0,0}1}%
}}}}
\put(3376,-7126){\makebox(0,0)[lb]{\smash{{\SetFigFont{17}{20.4}{\rmdefault}{\mddefault}{\updefault}{\color[rgb]{0,0,0}3}%
}}}}
\put(8641,-1276){\makebox(0,0)[lb]{\smash{{\SetFigFont{20}{24.0}{\rmdefault}{\mddefault}{\updefault}{\color[rgb]{.5,.17,0}Edge List (or $eList$)}%
}}}}
\put(1801,-5866){\rotatebox{90.0}{\makebox(0,0)[lb]{\smash{{\SetFigFont{20}{24.0}{\rmdefault}{\mddefault}{\updefault}{\color[rgb]{0,.56,0}Vertex List (or $vList$)}%
}}}}}
\end{picture}%

%% file: results.tex
\section{EXPERIMENTAL EVALUATION}
For the experiment, we consider a set of benchmarks generated for Ballot, Simple Auction, and Coin contracts from Solidity documentation \cite{Solidity}. Experiments are performed by varying the number of atomic-units, and threads. The analysis focuses on two main objectives: (1) Evaluate and analyzes the speedup achieved by concurrent miner over the serial miner. (2) Appraise the speedup achieved by concurrent validator over serial validator on various experiments.

\noindent \textbf{Experimental system:} The Experimental system is a large-scale 2-socket Intel(R) Xeon(R) CPU E5-2690 v4 @ 2.60 GHz with 14 cores per socket and two hyper-threads (HTs) per core, for a total of 56 threads. The machine has 32GB of RAM and runs Ubuntu 16.04.2 LTS.


\noindent \textbf{Methodology:} We have considered two types of workload, (W1) The number of atomic-units varies from 50 to 400, while threads and shared data-objects are fixed to 50 and 40 respectively. (W2) The number of threads varies from 10 to 60 while atomic-units are fixed to 400 and shared data-objects to 40. In all the experiments time taken by miners and validators is collected as an average of ten executions for the final result. 
\vspace{-.5cm}
\subsection{Benchmarks}
In reality, miner forms a block which consists of a set of transactions from different contracts. So, we consider four benchmarks Ballot, Simple Auction, Coin including Mixed contract which is the combination of above three. In Ethereum blockchain, smart contracts are written in Solidity and runs on the Ethereum Virtual Machine (EVM). The issue with EVM is that it does not support multi-threading and hence give poor throughput. Therefore, to exploit the efficient utilization of multi-core resources and to improve the performance, we convert smart contract from Solidity language into C++ and execute them using multi-threading. 
The details of the benchmarks are as follows:
\begin{enumerate}
    \item \textbf{Simple Auction:} It is an auction contract in which \textit{bidders, highest bidder}, and \textit{highest bid} are the shared data-objects. A single owner initiates the auction after that bidders can bid in the auction. The termination condition for auction is the bidding period (or end time) initialized at the beginning of the auction. During bidding period multiple bidders initiate their bids with biding amount using \textit{bid()} method. At the end of the auction, a bidder with the highest amount will be successful, and respective bid amount is transferred to the beneficiary account. 
    Conflict can occur if at least two bidders are going to request for bidPlusOne() simultaneously. 
   
    \item \textbf{Coin:} It is the simplest form of a cryptocurrency in which accounts are the shared data-objects. All accounts are uniquely identified by Etherum addresses. Only the contract deployer known as minter will be able to generate the coins and initialize the accounts at the beginning. Anyone having an account can send coins to another account with the condition that they have sufficient coins in their account or can check their balance. In the initial state, minter initializes all the accounts with some coins. 
    Conflict can occur if at least two senders are transferring the amount into the same receiver account simultaneously or when one send() and getbalance() have an account in common.
    \item \textbf{Ballot: }
It implements an electronic voting contract in which voters and proposals are the shared data-objects. All the voters and proposals are already registered and have unique Ethereum address. At first, all the voters are given rights by the chairperson (or contract deployer) to participate in the ballot. Voters either cast their vote to the proposal of their choice or delegate vote to another voter whom they can trust using delegate(). A voter is allowed to delegate or vote once throughout the ballot. Conflict can occur if at least two voters are going to delegate their vote to the same voter or cast a vote to the same proposal simultaneously.  
Once the ballot period is over, the winner of the ballot is decided based on the maximum vote count.
    \item \textbf{Mixed:} In this benchmark, we have combined all the above benchmarks in equal proportions. Data conflicts occur when AUs of the same contract executed simultaneously, and operate on common shared data-objects. 
\end{enumerate}
In all the above contracts, conflicts can very much transpire when \emph{miner} executes them concurrently. So, we use Optimistic STMs to ensure consistency and handle the conflicts. 
 \vspace{-.2cm}
\subsection{Results}

\noindent
We have shown the speedup of concurrent execution by miner and validator over serial in Table I. The results from the serial execution of the miner and validator are served as the baseline. 

 \begin{figure}
 	\centering
 	{\includegraphics[width=9cm]{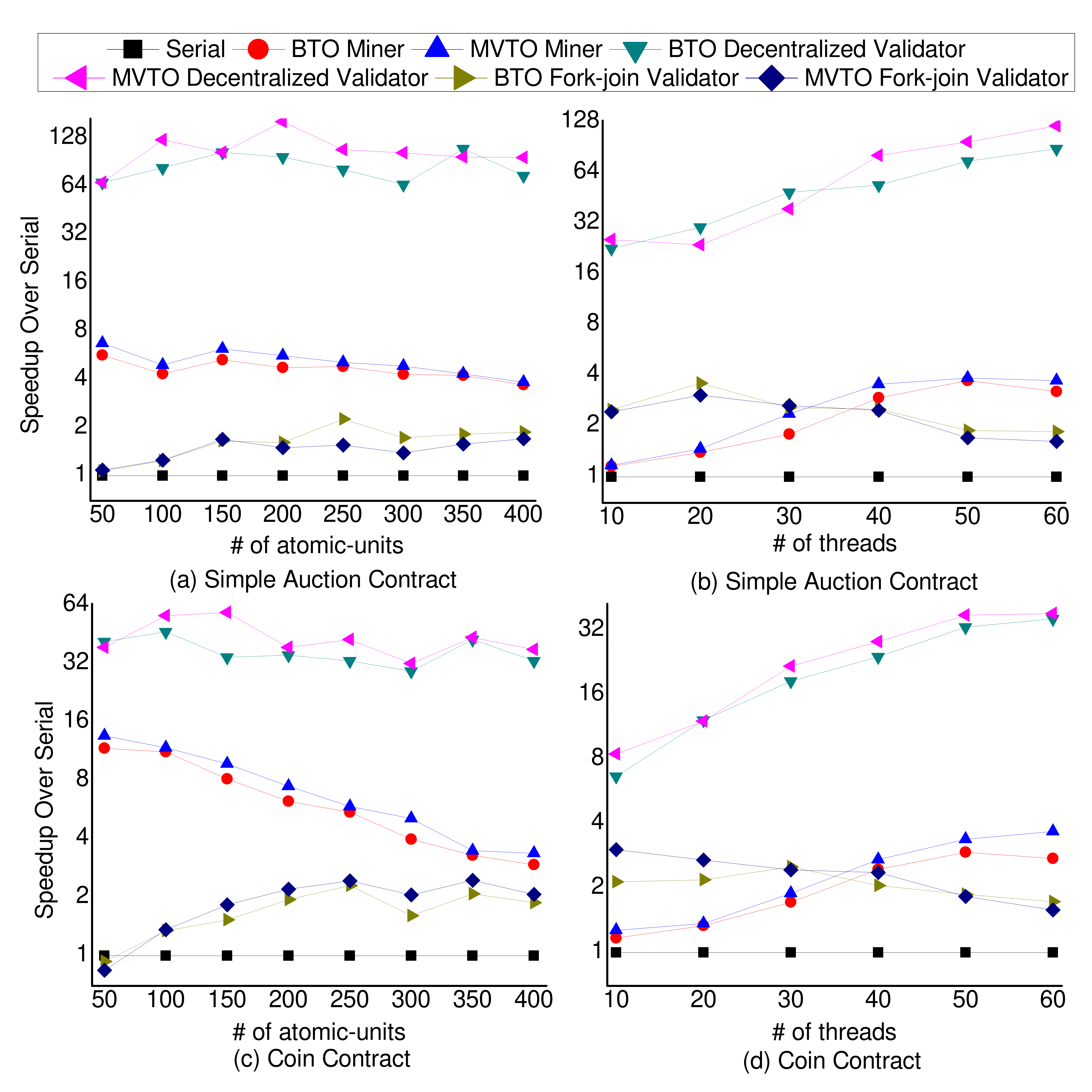}}
\vspace{-.8cm} 	\caption{Simple Auction and Coin Contracts}
 	\label{fig:asc}
 \end{figure}
\begin{figure}
	\centering
	{\includegraphics[width=9cm]{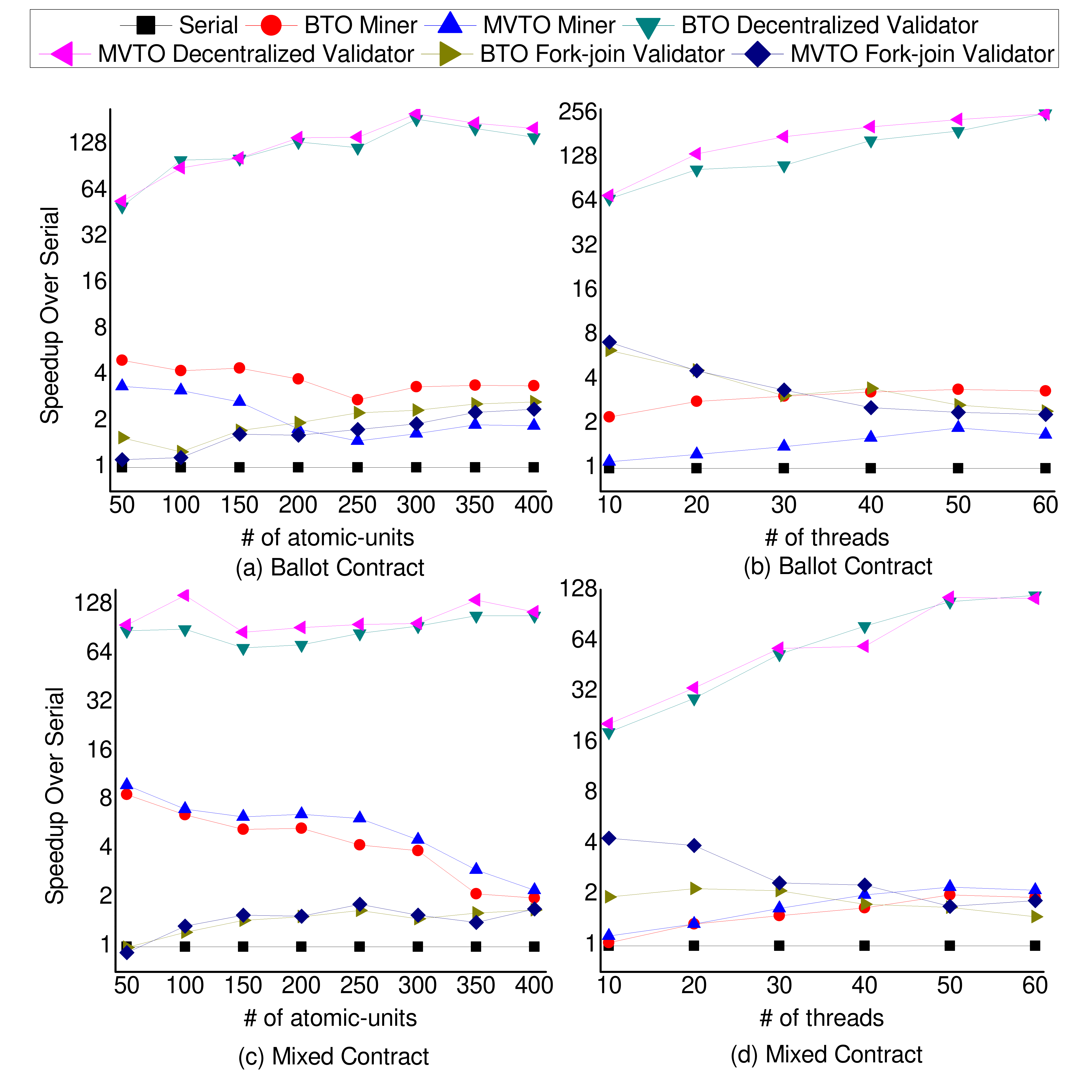}}
\vspace{-.7cm}	\caption{Ballot and Mixed Contracts}
	\label{fig:bmc}
\end{figure}

\figref{asc} and \figref{bmc} represent the speedup of concurrent miner and validator relative to the serial miner and validator for all the smart contracts on workload W1 and W2. It shows average speedup of 3.6x and 3.7x by the BTO and MVTO concurrent miner over serial miner respectively. Along with, BTO and MVTO validator outperforms average 40.8x and 47.1x than serial validator respectively \footnote{Code is available here: https://github.com/pdcrl/Blockchain}. The maximum speedup by concurrent miner on workload W1 is achieved at the smaller number of atomic-units. On workload W2 speedup of concurrent miner increases while increasing the number of threads up to fix number depending on system configuration.

The time taken by the concurrent validator is negligible as compared to serial validator because concurrent validator executes contracts concurrently and deterministically using BG given by concurrent miner. BG simplifies the parallelization task for the validator as validator need not to determine the conflicts, and directly executes non-conflicting transactions concurrently. 
It is clear from \figref{asc} and \figref{bmc} that BTO and MVTO Decentralized Validator is giving far better performance than BTO and MVTO Fork-join Validator. A possible reason can be master thread of BTO and MVTO Fork-join Validator becomes slow to assign the task to slave threads. 
\cmnt{
\begin{table*}[ht] \centering
	\begin{tabular}{|c|l|l|l|}
		\hline
		\multicolumn{1}{|c|}{Simple Auction} & \multicolumn{1}{c|}{Coin} & \multicolumn{1}{c|}{Ballot} & \multicolumn{1}{c|}{Mixed}\\
		\hline
		&  &  &\\
		\hline
		&  &  &\\
		\hline
		&  &  &\\
		\hline
		&  &  &\\
		\hline
		&  &  &\\
		\hline
		&  &  &\\
		\hline
	\end{tabular}
	\caption{Average speedups for each benchmark}
	\label{tbl:speedup}
\end{table*}
}
\begin{table}
	\caption{Speedup achieved by concurrent Miner and Validator}
	\vspace{-.3cm}
	\label{tbl:speedup}
	\resizebox{.49\textwidth}{!}{%
		\begin{tabular}{|c|c|c|c|c|c|c|c|c|}
			\hline
			\multirow{2}{*}{} & \multicolumn{2}{c|}{\textbf{Simple Auction}} & \multicolumn{2}{c|}{\textbf{Coin}} & \multicolumn{2}{c|}{\textbf{Ballot}} & \multicolumn{2}{c|}{\textbf{Mixed}} \\ \cline{2-9} 
			& \textbf{W1} & \textbf{W2} & \textbf{W1} & \textbf{W2} & \textbf{W1} & \textbf{W2} & \textbf{W1} & \textbf{W2} \\ \hline
			\textbf{BTO Miner} & 4.6 & 2.4 & 6.6 & 2.1 & 3.8 & 3.1 & 4.8 & 1.6 \\ \hline
			\textbf{MVTO Miner} & 5.2 & 2.7 & 7.5 & 2.4 & 2.3 & 1.5 & 5.7 & 1.8 \\ \hline
			\textbf{BTO Decentralized Validator} & 85.7 & 53.1 & 36.9 & 21.7 & 126.7 & 152.1 & 90.7 & 68.6 \\ \hline
			\textbf{MVTO Decentralized Validator} & 108.5 & 64.6 & 43.5 & 24.4 & 135.8 & 180.8 & 109.5 & 67.4 \\ \hline
			\textbf{BTO Fork-join Validator} & 1.7 & 2.5 & 1.7 & 2.1 & 2.1 & 3.8 & 1.5 & 1.9 \\ \hline
			\textbf{MVTO Fork-join Validator} & 1.5 & 2.3 & 1.9 & 2.3 & 1.8 & 3.8 & 1.5 & 2.7 \\ \hline
		\end{tabular}%
	}
\end{table}
\figref{asc} shows the speedup achieved by concurrent MVTO  Miner is greater than BTO Miner for Simple Auction and Coin contract on workload W1 and W2 respectively. A plausible reason can be that MVTO gives good performance for read-intensive workloads \cite{Kumar+:MVTO:ICDCN:2014}. Here, Simple Auction and Coin contracts are read-intensive \cite{Solidity}. 
 \figref{asc} (c) represents the speedup achieved by BTO and MVTO Fork-join Validators are even less than serial for 50 AUs due to the overhead of allocating the task by master thread. 

\figref{bmc} (a) and (b) capture better speedup achieved by concurrent BTO  Miner as compare to concurrent MVTO Miner for workload W1 and W2 because Ballot contract is write-intensive \cite{Solidity}. \figref{bmc} (c) and (d) represent the speedup achieved by concurrent miner and validator over serial miner and validator for the Mixed contract on workload W1 and W2 respectively. Due to equal proportions of all the above three contracts, the Mixed contract becomes read-intensive. So, the properties of the Mixed contract are same as Simple Auction and Coin contract with similar reasoning. Due to space constraints, we present essential results in the main paper and the remaining results on different workloads 
are available in the \apnref{rresult}.

%% file: conclusion.tex
\vspace{-.3cm}
\section{Conclusion}
\label{sec:con}
To exploit the multi-core processors, we have proposed the concurrent execution of \scontract{} by miners and validators which improves the throughput. Initially, miner executes the smart contracts concurrently using optimistic STM protocol as BTO. To reduce the number of aborts and improves the efficiency further, the concurrent miner uses MVTO protocol which maintains multiple versions corresponding to each data-object. 
Concurrent miner proposes a block which consists of a set of transactions, BG, hash of the previous block and final state of each shared data-objects. Later, the validators re-execute the same \SContract{} transactions concurrently and deterministically with the help of BG given by miner which capture the conflicting relations among the transactions to verify final state. If the validation is successful then proposed block appended into the blockchain and miner gets incentive otherwise discard the proposed block. Overall, BTO and MVTO miner performs
3.6x and 3.7x speedups over serial miner respectively. Along with, BTO and MVTO validator outperform average 40.8x and 47.1x than serial validator respectively. 

\vspace{1mm}
\noindent
\textbf{Acknowledgements.} This project is in part supported by a research grant from Thynkblynk Technologies Pvt. Ltd, and IMPRINT India scheme. We are grateful to Dr. Bapi Chatterjee, Dr. Sandeep Hans, Mr. Ajay Singh for several useful discussions that we had on this topic. We would like to thank anonymous reviewers for their useful comments. We are also very grateful to Anila Kumari  and G Monika the developers of IITH STM. 

%% file: appendix.tex
\section*{Appendix}
\label{apn:appendix}

\section{Detailed Experimental Evaluation}
\label{apn:rresult}
\subsection{Remaining Results}
\begin{figure}[H]
        	\centering
        	{\includegraphics[width=9cm]{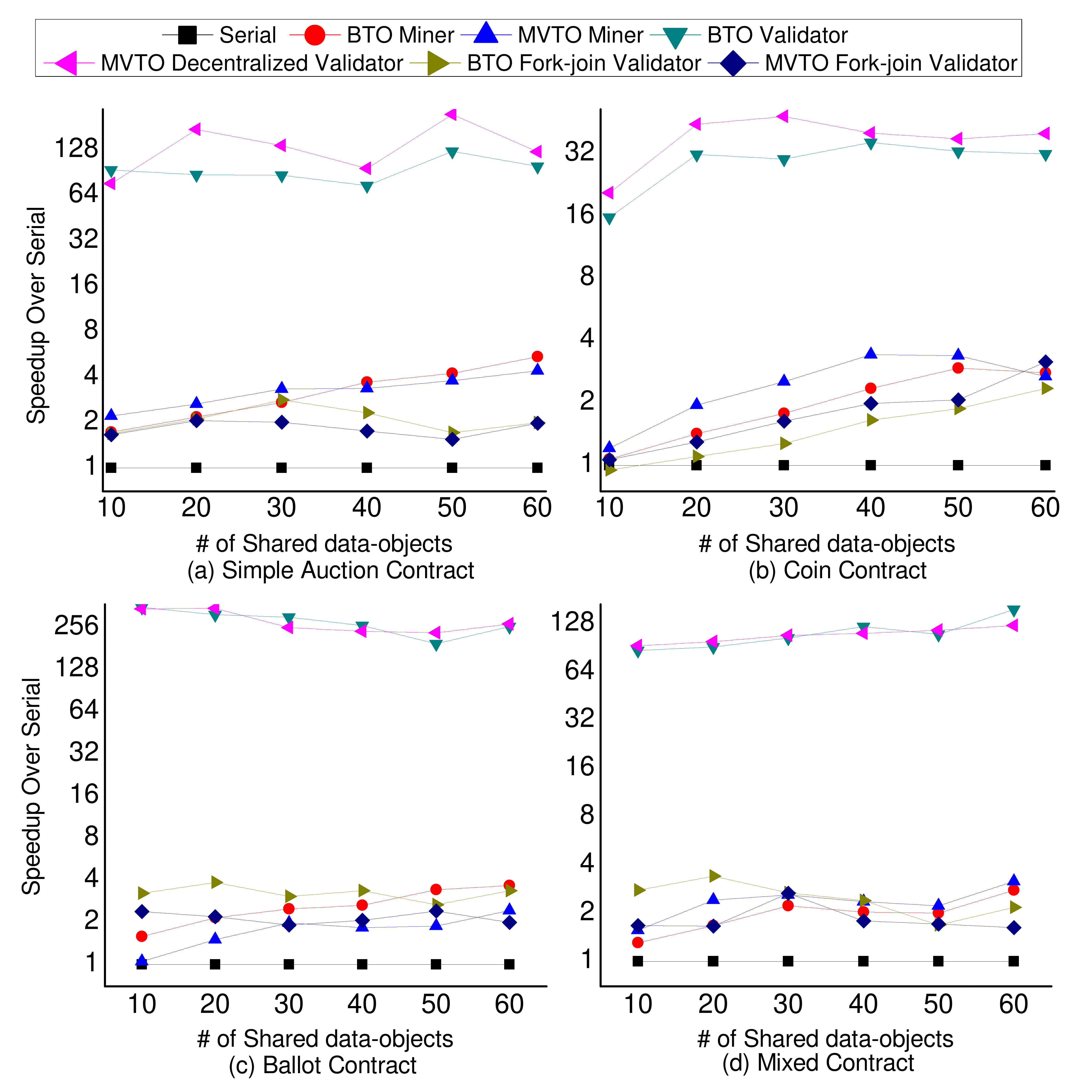}}
        	\caption{Speedup on Workload W3 for all Contracts}
        	\label{fig:allc}
        \end{figure}

In addition to workload W1 and W2, we analyze one more workload W3, which shows the result while varying the number of the shared data-objects from 10 to 60 and by fixing the number of threads and atomic-units to 50 and 400 respectively for all the contracts. As shown in \figref{allc}, it can be seen that with the increase in the number of shared data-objects speedup achieved by concurrent miner and validator also increases compared to the serial miner and validator. The reason behind this speedup is that data conflicts will decrease with the increase in shared data-objects, and a higher number of atomic-units can be executed concurrently.

Table II represents the speedup of concurrent miner and validator over serial miner and validator on various workloads W1, W2, and W3 for each contract.

\cmnt{
\begin{table}[]
	\caption{Speedup achieved by concurrent Miner and Validator over serial Miner and Validator.}
	\label{tbl:speedup}
	\resizebox{.5\textwidth}{!}{%
		\begin{tabular}{|c|c|c|c|c|c|c|c|c|}
			\hline
			\multirow{2}{*}{} & \multicolumn{2}{c|}{\textbf{Simple Auction}} & \multicolumn{2}{c|}{\textbf{Coin}} & \multicolumn{2}{c|}{\textbf{Ballot}} & \multicolumn{2}{c|}{\textbf{Mixed}} \\ \cline{2-13} 
			& \textbf{W1} & \textbf{W2} & \textbf{W1} & \textbf{W2} & \textbf{W1} & \textbf{W2}  & \textbf{W1} & \textbf{W2} \\ \hline
			\textbf{BTO Miner} & 4.6 & 2.4 & 6.6 & 2.1 & 3.8 & 3.1 & 4.8 & 1.6 \\ \hline
			\textbf{MVTO Miner} & 5.2 & 2.7 & 7.5 & 2.4 & 2.3 & 1.5 & 5.7 & 1.8 \\ \hline
			\textbf{BTO Decentralized Validator} & 85.7 & 53.1 & 36.9 & 21.7 & 126.7 & 152.1 & 90.7 & 68.6 \\ \hline
			\textbf{MVTO Decentralized Validator} & 108.5 & 64.6 & 43.5 & 24.4 & 135.8 & 180.8 & 109.5 & 67.4 \\ \hline
			\textbf{BTO Fork-join Validator} & 1.7 & 2.5 & 1.7 & 2.1 & 2.1 & 3.8 & 1.5 & 1.9 \\ \hline
			\textbf{MVTO Fork-join Validator} & 1.5 & 2.3 & 1.9 & 2.3 & 1.8 & 3.8 & 1.5 & 2.7 \\ \hline
		\end{tabular}%
	}
\end{table}
}

\subsection{Average Time taken by each Contract}    
\begin{figure}
	\centering
	{\includegraphics[width=9cm]{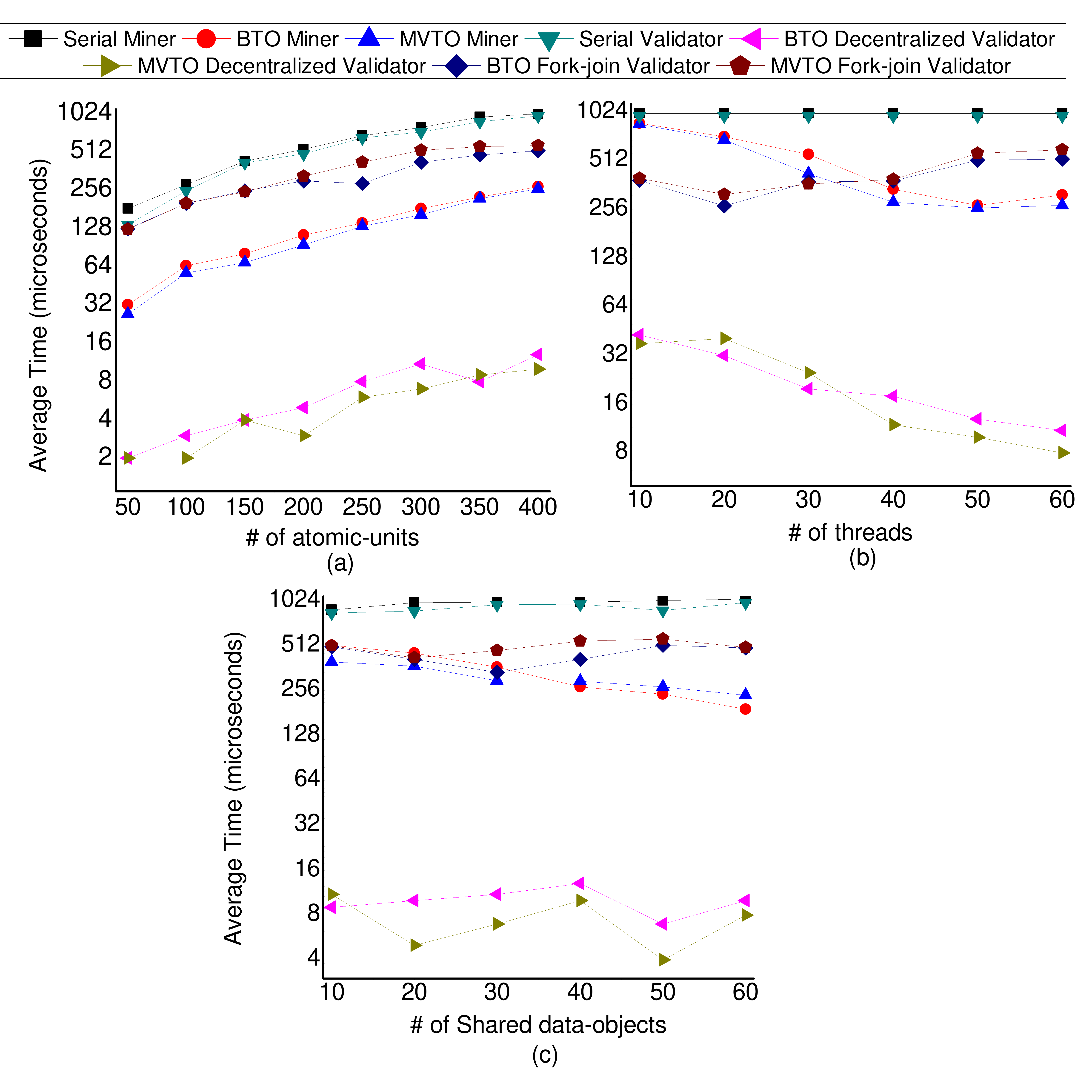}}
	\caption{Average Time taken by Simple Auction Contract}
	\label{fig:sa}
\end{figure}

\begin{figure}
	\centering
	{\includegraphics[width=9cm]{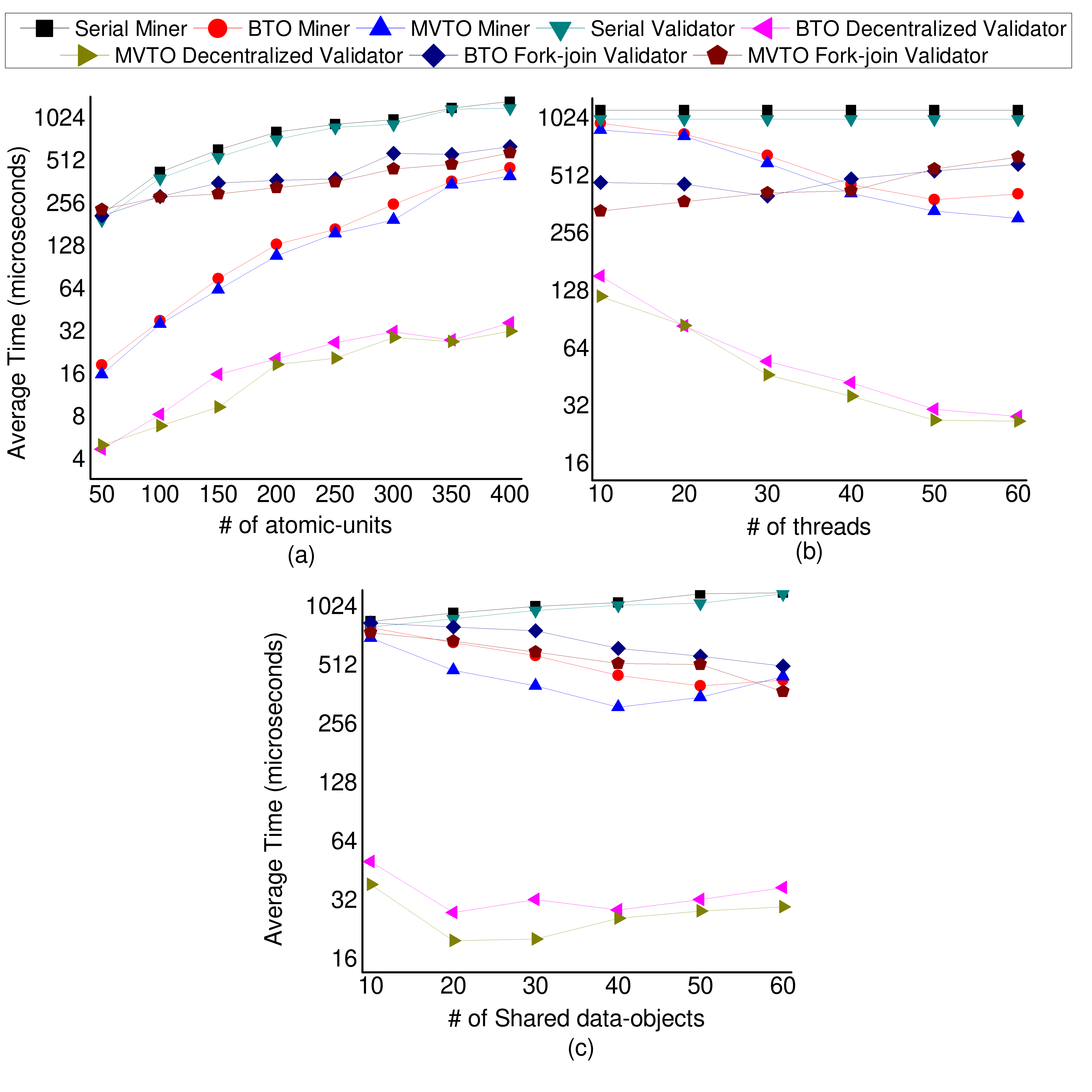}}
	\caption{Average Time taken by Coin Contract}
	\label{fig:coin}
\end{figure}

\begin{figure}
	\centering
	{\includegraphics[width=9cm]{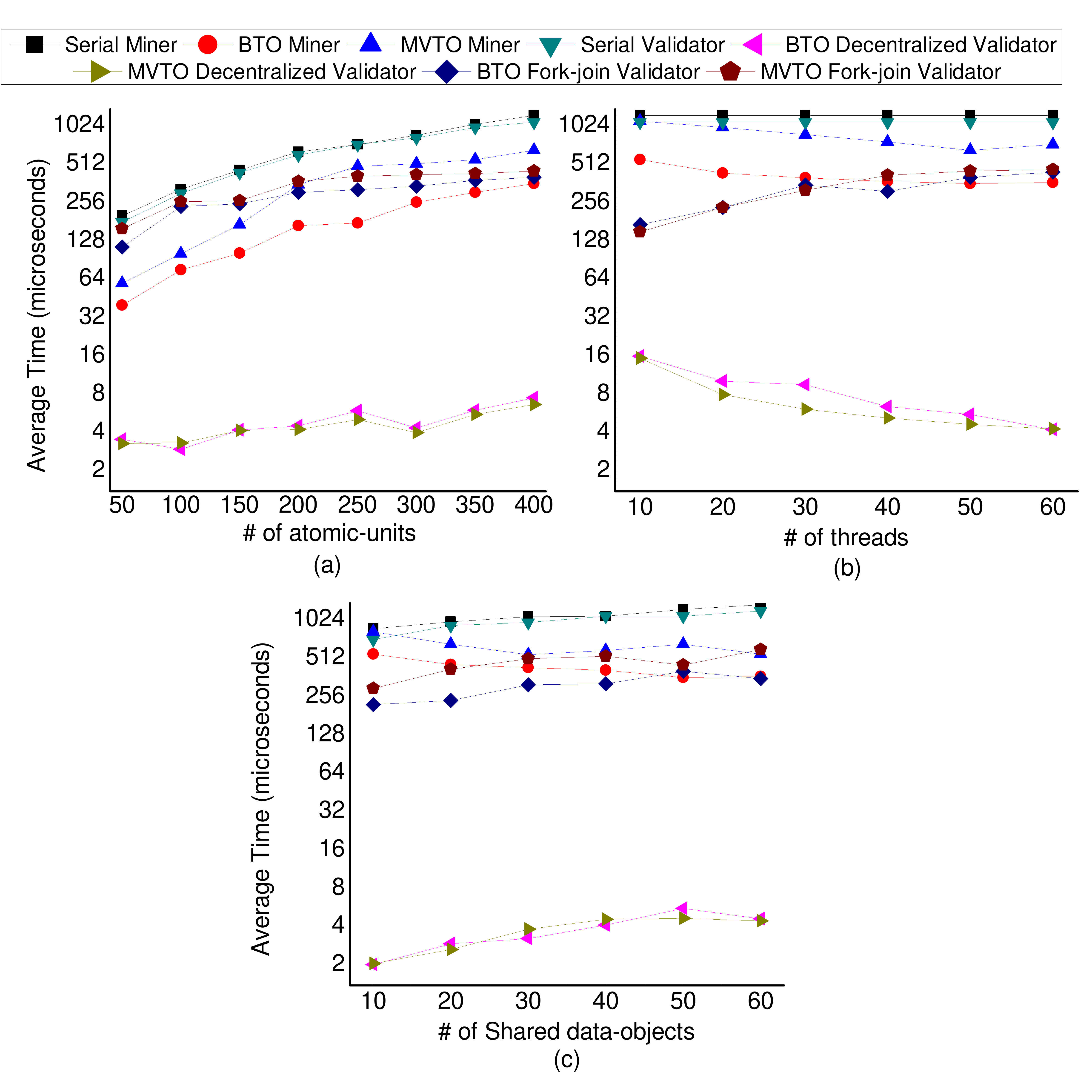}}
	\caption{Average Time taken by Ballot Contract}
	\label{fig:ballotc}
\end{figure}

\begin{figure}[H]
	\centering
	{\includegraphics[width=9cm]{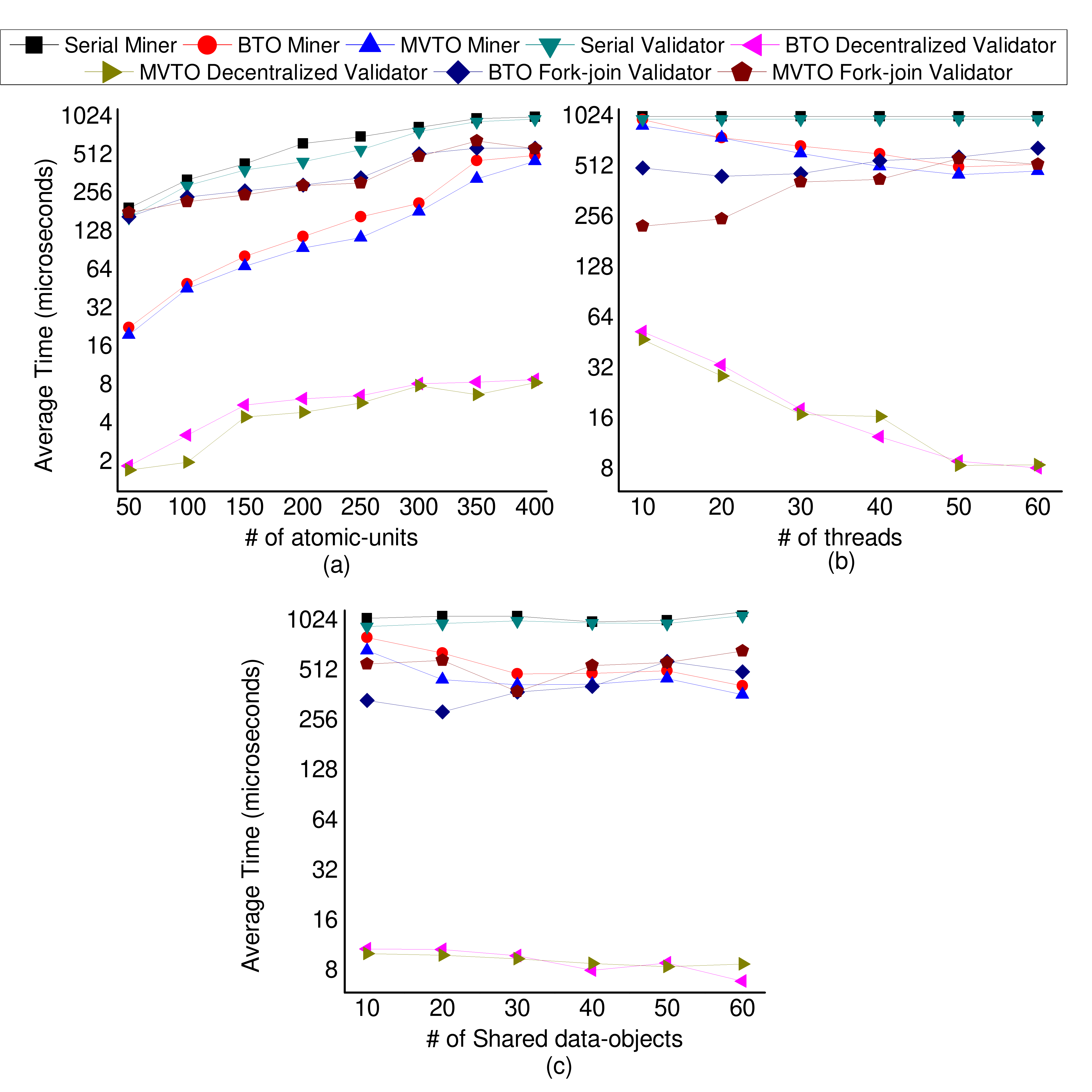}}
	\caption{Average Time taken by Mixed Contract}
	\label{fig:mix}
\end{figure}

\figref{sa}, \figref{coin}, \figref{ballotc} and \figref{mix}, represents the average time taken by Simple Auction, Coin, Ballot and Mixed contracts benchmark respectively. It can be seen that time taken by serial miner and validator is higher than the proposed concurrent miner and validator. Moreover, the serial validator is taking less time than the serial miner this is because validator will only get the valid transaction in the block given by the miner. From \figref{ballotc}, this can be observed that for write-intensive workload performance of BTO Miner is better than MVTO Miner. However, for all other workloads which are read-intensive MVTO Miner gives better performance. 

For all the benchmarks, the concurrent validator is taking less time compared to the concurrent miner because concurrent miner did all the required task to find the data conflict and generates the BG to help the validator. In this process, the validator will get a deterministic order of execution in the form of BG and without bothering about data conflicts, validator executes the atomic-units concurrently. However, it can be seen in all the figures, BTO and MVTO Decentralized Validator dominates a BTO and MVTO Fork-join Validator because of overhead associated with a master thread to assign the tasks (atomic-units) to hungry slave threads.

\section{Pcode of Lock-free Graph Library}
\label{apn:rpcode}

\subsection{Lock-free Concurrent Block Graph}
\label{apn:apnbg}

\noindent
\textbf{Lock-free Graph Library Methods Accessed by Concurrent Miner:} Concurrent miner uses addVert() and addEdge() methods of lock-free graph library to build a block graph. When concurrent miner wants to add a node in the block graph then first it calls addVert() method. addVert() method identifies the correct location of that node (or \vgn{}) in the \vl{} at \Lineref{addv2}. If \vgn{} is not part of \vl{} then it creates the node and adds it into \vl{} at \Lineref{addv5} in lock-free manner with the help of atomic compare and swap operation. Otherwise, \vgn{} is already present in \vl{} at \Lineref{addv10}.

After successful addition of \vnode{} in the block graph concurrent miner calls addEdge() method to add the conflicting node (or \egn{}) corresponding to \vnode{} in the \el{}. First, addEdge() method identifies the correct location of \egn{} in the \el{} of corresponding \vnode{} at \Lineref{adde4}. If \egn{} is not part of \el{} then it creates the node and adds it into \el{} of \vnode{} at \Lineref{adde7} in lock-free manner with the help of atomic compare and swap operation. After successful addition of \enode{} in the \el{} of \vnode{}, it increment the \inc{} of \enode.$vref$ (to maintain the indegree count) node which is present in the \vl{} at \Lineref{adde8}.

\begin{algorithm}[H]
	\scriptsize
	\label{alg:cg} 
	\caption{BG(\emph{vNode}, STM): It generates a block graph for all the atomic-unit nodes.}
	\begin{algorithmic}[1]
		\makeatletter\setcounter{ALG@line}{72}\makeatother
		\Procedure{BG(\emph{vNode}, STM)}{} \label{lin:cg1}
		\State /*Get the \cl{} of committed transaction $T_i$ from STM*/\label{lin:cg2}
		\State clist $\gets$ STM.\gconfl(\emph{vNode}.$ts_i$);\label{lin:cg3}
		\State /*Transaction $T_i$ conflicts with $T_j$ and $T_j$ existes in conflict list of $T_i$*/\label{lin:cg4}
		\ForAll{($ts_j$ $\in$ clist)}\label{lin:cg5}
		\State \addv(\emph{$ts_j$}); \label{lin:cg6}
		\State \addv(\emph{vNode.$ts_i$});\label{lin:cg7}
		\If{($ts_j$  $<$ \emph{vNode}.$ts_i$)}\label{lin:cg8}
		\State \adde($ts_j$, \emph{vNode}.$ts_i$);\label{lin:cg9}
		\Else\label{lin:cg10}
		\State \adde(\emph{vNode}.$ts_i$, $ts_j$);\label{lin:cg11}
		\EndIf  \label{lin:cg12}
		\EndFor\label{lin:cg13}
		\EndProcedure\label{lin:cg14}
	\end{algorithmic}
\end{algorithm}

\begin{algorithm}[H]
	\scriptsize
	\label{alg:addv} 	
	\caption{\addv{($ts_i$)}: It finds the appropriate location of vertex graph node (or \vgn{}) which is having a $ts$ as $i$ in the \vl{} and add into it. } 
	\begin{algorithmic}[1]
		\makeatletter\setcounter{ALG@line}{86}\makeatother
		\Procedure{\addv{($ts_i$)}}{} \label{lin:addv1}
		\State Identify the $\langle$\vp, \vc{}$\rangle$ of \vgn{} of $ts_i$ in \vl{} of $BG$;\label{lin:addv2}
		\If{(\vc.$ts_i$ $\neq$ \vgn.$ts_i$)}\label{lin:addv3}
		\State Create new Graph Node (or \vgn) of $ts_i$ in \vl{};\label{lin:addv4}
		\If{(\vp.\vn.CAS(\vc, \vgn))}\label{lin:addv5}
		
		\State return$\langle$\emph{Vertex added}$\rangle$; /*\vgn{} is successfully inserted in \vl{}*/ \label{lin:addv6}
		\EndIf\label{lin:addv7}
		\State goto \Lineref{addv2}; /*Start with the \vp{} to identify the new $\langle$\vp, \vc{}$\rangle$*/ \label{lin:addv8}
		\Else\label{lin:addv9}
		\State return$\langle$\emph{Vertex already present}$\rangle$; /*\vgn{} is already present in \vl{}*/\label{lin:addv10}
		\EndIf\label{lin:addv11}
		\EndProcedure\label{lin:addv12}
	\end{algorithmic}
\end{algorithm}

\begin{algorithm}[H]
	\scriptsize
	\label{alg:adde} 	
	\caption{\adde{\emph{(fromNode, toNode)}}: It adds an  edge from \emph{fromNode} to \emph{toNode}.}
	\begin{algorithmic}[1]
		\makeatletter\setcounter{ALG@line}{98}\makeatother
		\Procedure{\adde{\emph{(fromNode, toNode)}}}{}\label{lin:adde1}
		\State Identify the $\langle$\ep, \ec{}$\rangle$ of \emph{toNode} in \el{} of the \emph{fromNode} vertex in $BG$;\label{lin:adde4}
		\If{(\ec.$ts_i$ $\neq$ toNode.$ts_i$)}\label{lin:adde5}
		\State Create new Graph Node (or \egn) in \el{};\label{lin:adde6} 
		\If{(\ep.\en.CAS(\ec, \egn))}\label{lin:adde7}
		\State Increment the \inc{} atomically of \egn.\emph{vref} in \vl{};\label{lin:adde8}
		\State return$\langle$\emph{Edge added}$\rangle$; /*toNode is successfully inserted in \el{}*/ \label{lin:adde9}
		\EndIf\label{lin:adde10}
		\State goto \Lineref{adde4}; /*Start with the \ep{} to identify the new $\langle$\ep, \ec{}$\rangle$*/\label{lin:adde11}
		\Else\label{lin:adde12}
		\State return$\langle$\emph{Edge already present}$\rangle$; /*toNode is already present in \el{}*/\label{lin:adde13}
		\EndIf\label{lin:adde14}
		\EndProcedure\label{lin:adde15}
	\end{algorithmic}
\end{algorithm}

\noindent
\textbf{Lock-free Graph Library Methods Accessed by Concurrent Validator:} Concurrent validator uses searchLocal(), searchGlobal() and decInCount() methods of lock-free graph library. First, concurrent validator thread calls searchLocal() method to identify the source node (having indegree (or \inc) 0) in its local \cachel{} (or thread local memory). 
If any source node exist in the local \cachel{} with \inc{} 0 then it sets \inc{} field to be -1 at \Lineref{sl1} atomically. 

If source node does not exist in the local \cachel{} then concurrent validator thread calls searchGlobal() method to identify the source node in the block graph at \Lineref{sg1}. If any source node exists in the block graph then it will do the same process as done by searchLocal() method. After that validator thread calls the decInCount() method to decreases the \inc{} of all the conflicting nodes atomically which are present in the \el{} of corresponding source node at \Lineref{ren1}. While decrementing the \inc{} of each conflicting nodes in the block graph, it again checks if any conflicting node became a source node then it adds that node into its local \cachel{} to optimize the search time of identifying the next source node at \Lineref{ren3}.

\begin{algorithm}[H]
	\scriptsize
	\caption{\searchl{(cacheVer, $AU_{id}$)}: First validator thread search into its local \cachel{}.}
	\begin{algorithmic}[1]
		\makeatletter\setcounter{ALG@line}{111}\makeatother
		\Procedure{\searchl{($cacheVer$)}}{}
		\If{(cacheVer.\inc.CAS(0, -1))} \label{lin:sl1} 
		\State \nc{} $\gets$ \nc{}.$get\&Inc()$; \label{lin:sl2}
		\State $AU_{id}$ $\gets$ cacheVer.$AU_{id}$;
		\State return$\langle$cacheVer$\rangle$;\label{lin:sl5}
		\Else\label{lin:sl6}
		\State return$\langle nil \rangle$;\label{lin:sl7}
		\EndIf\label{lin:sl8}
		\EndProcedure
	\end{algorithmic}
\end{algorithm}

\begin{algorithm}[H]
	\scriptsize
	\caption{\searchg{(BG, $AU_{id}$)}: Search the source node in the block graph whose \inc{} is 0.}
	\begin{algorithmic}[1]
		\makeatletter\setcounter{ALG@line}{120}\makeatother
		\Procedure{\searchg{(BG, $AU_{id}$)}}{}
		\State \vnode{} $\gets$ BG.\vh;	
		\While{(\vnode.\vn{} $\neq$ BG.\vt)} /*Search into the Block Graph*/			
		\If{(\vnode.\inc.CAS(0, -1))}\label{lin:sg1}
		\State \nc{} $\gets$ \nc{}.$get\&Inc()$; \label{lin:sg2}
		\State $AU_{id}$ $\gets$ \vnode.$AU_{id}$;
		\State return$\langle \vnode \rangle$;\label{lin:sg5}
		\EndIf\label{lin:sg8}
		\State \vnode $\gets$ \vnode.\vn;	
		\EndWhile
		\State return$\langle nil \rangle$;\label{lin:sg7}
		\EndProcedure
	\end{algorithmic}
\end{algorithm}

\begin{algorithm}[H]
	\scriptsize
	\caption{decInCount(remNode): Decrement the \inc{} of each conflicting node of source node.}
	\begin{algorithmic}[1]
		\makeatletter\setcounter{ALG@line}{132}\makeatother
		\Procedure{\emph{decInCount(remNode)}}{}
		\While{(remNode.\en $\neq$ remNode.\et)}	
		\State Decrement the \emph{inCnt} atomically of remNode.\emph{vref} in the \vl{}; \label{lin:ren1}
		\If{(remNode.\emph{vref}.\inc{} == 0)}\label{lin:ren2}
		\State Add remNode.\emph{verf} node into \cachel{} of thread local log, \tl{};\label{lin:ren3}
		\EndIf\label{lin:ren4}
		\State remNode $\gets$ remNode.\en.\emph{verf};		
		\State return$\langle$remNode$\rangle$;
		\EndWhile
		\State return$\langle nil \rangle$;
		\EndProcedure		
	\end{algorithmic}
\end{algorithm}

\begin{algorithm} [H]
\scriptsize
	\label{alg:exec} 
	\caption{\exec{(curAU)}: Execute the current atomic-units.}
	\begin{algorithmic}[1]
		\makeatletter\setcounter{ALG@line}{143}\makeatother
		\Procedure{\exec{($curAU$)}}{}
		\While{(curAU.steps.hasNext())} /*Assume that curAU is a list of steps*/
		\State curStep = currAU.steps.next(); /*Get the next step to execute*/
		\Switch{(curStep)}
		\EndSwitch
		\Case{read($x$):}
		\State Read Shared data-object $x$ from a shared memory;
		\EndCase
		\Case{write($x, v$):} 
		\State Write Shared data-object $x$ in shared memory with value $v$;
		\EndCase
		\Case{default:}
		\State /*Neither read from or write to a shared memory Shared data-objects*/;
		\State execute curStep;
		\EndCase
		\EndWhile			
		\State return $\langle void \rangle$	
		\EndProcedure			
	\end{algorithmic}
\end{algorithm}

\section{Basic Timestamp Ordering (BTO) Algorithm}
\label{sec:bto}

We start with data-structures that are local to each transaction $T_i$ as its local log. For each transaction $T_i$:

\begin{itemize}
	\item $rset_i$(read-set): It is a list of data tuples ($d\_tuples$) of the form $\langle x, v \rangle$, where $x$ is the \tobj{} and $v$ is the value read by the transaction $T_i$. We refer to a tuple in $T_i$'s read-set by $rset_i[x]$.
	
	\item $wset_i$(write-set): It is a list of $d\_tuples$ of the form $\langle x, v \rangle$, where $x$ is the \tobj{} to which transaction $T_i$ writes the value $v$. Similarly, we refer to a tuple in $T_i$'s write-set by $wset_i[x]$.
	
\end{itemize}

\noindent In addition to these local structures, the following shared global structures are maintained that are shared across transactions (and hence, threads).

\begin{itemize}
	\item $\tcntr{}$ (counter): This is a numerical valued counter that is incremented atomically when a transaction begins.
\end{itemize}

\noindent For each \tobj{} $x$, we maintain:
\begin{itemize}
	
	\item $x.key$ (key): It represents the key corresponding to $x$.
	\item $x.v$ (value): It is a value $v$ written by a committed transaction.
	\item $x.lock()$: It acquires the lock on \tobj{} $x$. It is a boolean variable in which 0 represents $x$ is unlocked whereas 1 represents $x$ is locked.
	
	\item $\mr(x)$ (Maximum read): It is the timestamp of latest transaction that has read $x$. 
	
	\item $\mw(x)$ (Maximum write): It is the timestamp of latest transaction that has written into $x$.

	\item $x.rl$ (readList): It is the read list consists of all the transactions that have read $x$.
	
	\item $x.wl$ (writeList): It is the write list consists of all the transactions that written into $x$.  
	
\end{itemize}
\begin{algorithm} [H]
	\scriptsize
	\label{alg:init} 
	\caption{STM.\init{()}: Invoked at the start of the STM system; Initializes all the \tobj{s} used by the STM system.}
	\begin{algorithmic}[1]
		\makeatletter\setcounter{ALG@line}{157}\makeatother
		\Procedure{STM.\init{()}}{}
		\State \tcntr{} = 1; 
		\ForAll {$x$ used by the STM system}
		\State /* $T_0$ is initializing $x$ with $\langle$ \emph{key, v, lock, max\_r, max\_w, rl, wl}$\rangle$*/
		\State add $\langle$ \emph{x, 0, 0, 0, 0, null, null} $\rangle$ to \shl;/*\shl{} is a \emph{shared object list}*/ \label{lin:t0-init} 
		\EndFor
		\EndProcedure
	\end{algorithmic}
\end{algorithm}

\begin{algorithm} [H]
	\scriptsize
	\label{alg:begin} 
	\caption{STM.\begtrans{()}: Invoked by a thread to being a new transaction $T_i$.}
	\begin{algorithmic}[1]
		\makeatletter\setcounter{ALG@line}{164}\makeatother
		\Procedure{STM.\begtrans{()}}{}
		\State /*Store the latest value of $\tcntr$ in $i$;*/ 
		\State $i = \tcntr$; 
		\State \tcntr = \tcntr.$get\&Inc()$; 
		\State return $\langle i \rangle$; \label{lin:begfin}
		\EndProcedure
	\end{algorithmic}
\end{algorithm}

\begin{algorithm}  [H]
	\scriptsize
	\label{alg:read} 
	\caption{STM.\readi{(x)}: A transaction $T_i$ reads \tobj{} $x$.}
	\begin{algorithmic}[1]
		\makeatletter\setcounter{ALG@line}{170}\makeatother
		\Procedure{STM.\readi{(x)}}{}
		\State /*Check if the \tobj{} $x$ is in 
		$wset_i$ or $rset_i$*/
		\If {($x \in wset_i$)} 
		\State return $\langle wset_i[x].v \rangle$;
		\ElsIf {($x \in rset_i$)} 
		\State return $\langle rset_i[x].v \rangle$;
		\Else
		\State /*\tobj{} $x$ is not in $wset_i$ and $rset_i$*/ 
		\State $x.lock()$;
		\label{lin:rlock}
		\State /*Get the \mr{} and \mw{} timestamps of the $x$*/ 
		\If {(\mw(x) $> i$)} 
		\State return $\langle abort(i) \rangle$; 
		\EndIf
		
		\If {(\mr(x) $< i$)} 
		\State \mr(x) = $i$;
		\EndIf	
		\State $x.unlock()$;
		\State Append the $d\_tuple \langle x \rangle$ to $rset_i$ of $T_i$;
		\State return $\langle v \rangle$; /* return the value as $v$*/
		\EndIf
		\EndProcedure		
	\end{algorithmic}
\end{algorithm}

\cmnt{
	\begin{algorithm}  [H]
		\label{alg:select}
		\caption{$\cl($i$)$: Maintain the \cl{} corresponding to $T_i$;}
		\begin{algorithmic}[1]
			\State Append the $ts$ into \cl{} os $i$;
		\end{algorithmic}
	\end{algorithm}
}

\begin{algorithm} [H]  
	\scriptsize
	\label{alg:write} 
	\caption{STM.$write_i(x,v)$: A Transaction $T_i$ writes into local memory.}
	\begin{algorithmic}[1]
		\makeatletter\setcounter{ALG@line}{191}\makeatother
		\Procedure{STM.$write_i(x,v)$}{}		
		\If {($x \in wset_i$)} 
		\State Update the $d\_tuple \langle x,v \rangle$ to $wset_i$ of $T_i$;
		\Else
		\State Append the $d\_tuple \langle x,v \rangle$ to $wset_i$ of $T_i$;
		\EndIf
		\State return $\langle ok \rangle$;
		\EndProcedure
	\end{algorithmic}
\end{algorithm}

\begin{algorithm} [H] 
	\scriptsize
	\label{alg:tryc} 
	\caption{\tryc{$_i$(}): Returns $ok$ on \emph{commit} else return \emph{abort}.}
	\begin{algorithmic}[1]
		\makeatletter\setcounter{ALG@line}{199}\makeatother
		\Procedure{\tryc{$_i$(})}{}
		\ForAll {$d\_tuple (x,v)$ in $wset_i$}
		\State /* Lock the \tobj{s} in a predefined order to avoid deadlocks */
		\State $x.lock()$;
		
		\If {((\mw(x) $> i$) $||$ (\mr(x) $> i$))} 
		\State return $\langle abort(i) \rangle$; 
		\EndIf
		\EndFor
		\State /* Successfully done with the validation of $wset_i$ and not yet aborted; So the new write value should be updated; */
		\ForAll {$d\_tuples \langle x,v \rangle $ in $wset_i$}
		\State /*Set the \emph{(x.v = v)} and (\mw\emph{{(x)} = i})*/
		\State Update the \tobj{} $x$ with $\langle$ \emph{key, v, 1, max\_r, i, rl, wl}$\rangle$ in \shl;
		\State /*\cl{$[i]$} contains the all the conflicting transactions of $T_i$*/
		\State Insert the $x.rl$ and $x.wl$ into \cl{$[i]$};
		\State Add $i$ into the $x.wl$;
		\State $x.unlock()$;
		\EndFor
		\ForAll {\tobj{} $x$ in $rset_i$}
		\State /* Lock the \tobj{s} in a predefined order to avoid deadlocks; */
		\State $x.lock()$;
		\State Insert the $x.wl$ into \cl$[i]$;
		\State Add $i$ into the $x.rl$;
		\State $x.unlock()$;
		\EndFor
		\State return $\langle ok \rangle$;
		\EndProcedure
	\end{algorithmic}
\end{algorithm}

\begin{algorithm}[H] 
	\scriptsize
	\label{alg:ap-abort} 
	\caption{$abort(i)$: Invoked by various STM methods to abort transaction $T_i$. It returns $\mathcal{A}$.}
	\begin{algorithmic}[1]
		\makeatletter\setcounter{ALG@line}{225}\makeatother
		\Procedure{$abort(i)$}{}
		\State unlock all \tobj{s} locked by $T_i$ and release its local log;      
		\State return $\langle \mathcal{A} \rangle$;
		\EndProcedure
	\end{algorithmic}
\end{algorithm}

\begin{algorithm}[H] 
\scriptsize
	\label{alg:getconfl} 
	\caption{STM.\gconfl{($i$)}: Get the conflict list corresponding to transaction $T_i$}
	\begin{algorithmic}[1]
		\makeatletter\setcounter{ALG@line}{229}\makeatother
		\Procedure{STM.\gconfl{($i$)}}{}
		\State Transaction $T_i$ search the respective conflict in conflict list (or \cl{}).	
		\State return $\langle \cl$[i]$ \rangle$;
		\EndProcedure
	\end{algorithmic}
\end{algorithm}

\section{Multiversion Timestamp Ordering (MVTO) Algorithm}
\label{sec:mvto}

Local data structures of MVTO for each transaction $T_i$ is same as BTO local data structures. For each transaction $T_i$ it maintains $rset_i$ and $wset_i$ locally.
\cmnt{
\begin{itemize}
	\item $rset_i$(read-set): It is a list of data tuples ($d\_tuples$) of the form $\langle x, v \rangle$, where $x$ is the \tobj{} and $v$ is the value read by the transaction $T_i$. We refer to a tuple in $T_i$'s read-set by $rset_i[x]$.
	
	\item $wset_i$(write-set): It is a list of $d\_tuples$ of the form $\langle x, v \rangle$, where $x$ is the \tobj{} to which transaction $T_i$ writes the value $v$. Similarly, we refer to a tuple in $T_i$'s write-set by $wset_i[x]$.
	
\end{itemize}
}
\noindent In addition to these local structures, the following shared global structures are maintained that are shared across transactions (and hence, threads). For each transaction object (\tobj{}) $x$:
\begin{itemize}
	\item $x.vl$(version list): It is a list consisting of version tuples ($v\_tuple$) of the form $\langle ts, v, max_r, rl, \vn \rangle$ where $ts$ is the timestamp of the committed transaction that writes the value $v$ to $x$. $max_r$ contains the timestamp of latest transaction that has read $x$. The list $rl$ is the read list consisting of a set of transactions that have read the value $v$ (described below). Informally the version list consists of all the committed transaction that have ever written to this t-object and the set of corresponding transactions that have read each of those values. \vn{} will points to the next version in version list.
	
	\item $rl$(read list): This list contains all the read transaction tuples ($rt\_tuples$) of the form $\langle j \rangle$, where $j$ is the timestamp of the $commited$ reading transaction. The read list $rl$ is stored in each tuple of the version list.
\end{itemize}
\cmnt{
	\begin{figure} [tbph]
		\centerline{\scalebox{0.5}{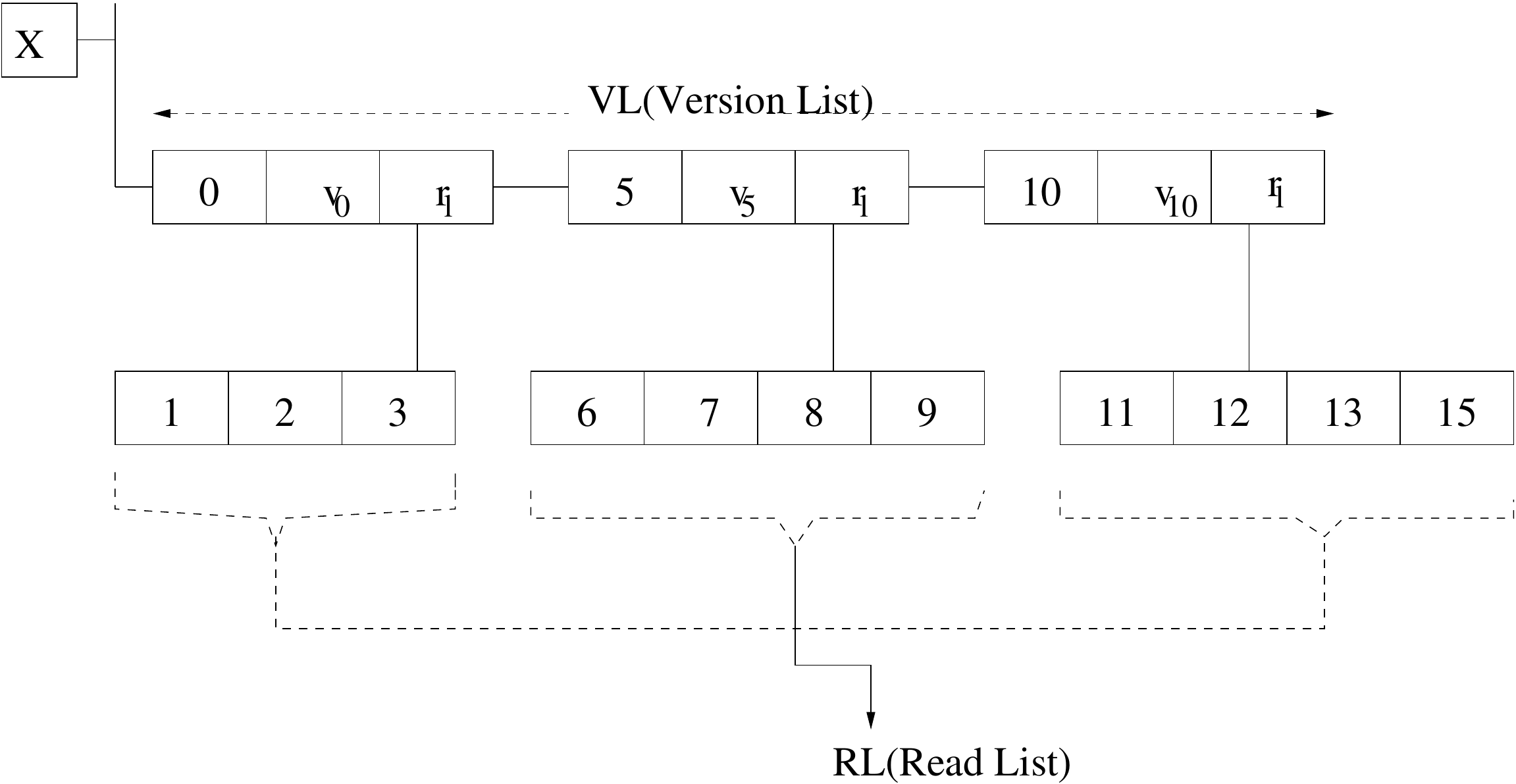}}
		\caption{Data Structures for Maintaining Versions}
		\label{fig:datas}
	\end{figure}
	
}

\begin{itemize}
	\item \tcntr: This counter is atomic and used to generate the ids/timestamp for a newly invoked transaction as BTO. This is incremented everytime a new transaction is invoked.
	\item \cl{$[i]$} (\emph{Conflict list}): It contains all the conflicts of transactions $T_i$.
	
\end{itemize}
\cmnt{
\begin{figure}
	\centerline{\scalebox{0.35}{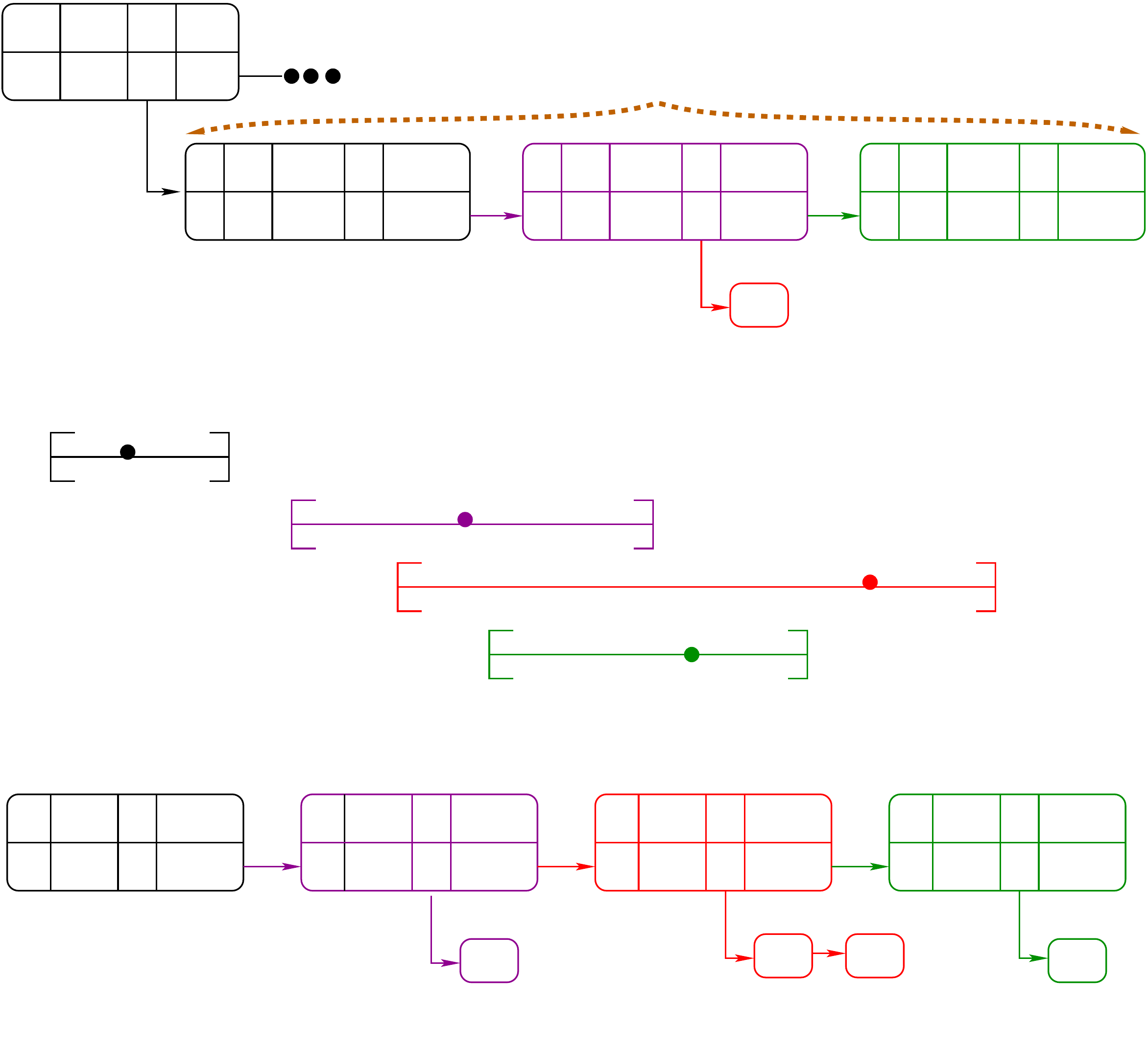}}
	\caption{Execution of MVTO}
	\label{fig:confg}
\end{figure}
}

\cmnt{
	%
	The STM system consists of the following operations/functions. These are executed whenever a transaction begins, reads, write or tries to commit:
	
	\vspace{1mm}
	\noindent
	\textit{$\init():$} This operation initializes the STM system. It is assumed that the STM system knows all the \tobj{s} ever accessed. All these \tobj{s} are initialized with value 0 by the initial transaction $T_0$ in this operation. A version tuple $\langle 0, 0, nil \rangle$ is inserted into all the version list of all the \tobj{s}. 
	
	\vspace{1mm}
	\noindent
	\textit{$\begtrans():$} A thread invokes a transaction by executing this operation. It returns an unique transaction identifier which is also its timestamp. The id is used in all other operations exported by the STM system. The id is further stored in the $\livel$.
	
	\vspace{1mm}
	\noindent
	\textit{$read_i(x):$} This operation is invoked when transaction $T_i$ wants to read a \tobj{} $x$. First, the \tobj{} $x$ is locked. Then the version list of $x$ is searched to identify the correct $version\_tuple$ (i.e the version created by a writing transaction). The tuple with the largest timestamp  less than $i$, say $\langle j, v \rangle$ is identified from the version-list. Then, $i$ is added to the read list of $j$'s version tuple. Finally, the value $v$ written by transaction $j$, is returned. 
	
	\vspace{1mm}
	\noindent
	\textit{$\find(i,x):$} This function is invoked by $read_i(x)$ and finds the tuple $\langle j,v,rl \rangle$ having the \underline{l}argest \underline{t}imestamp value $j$ \underline{s}maller than $i$ (denoted as lts). 
	
	\vspace{1mm}
	\noindent
	\textit{$write_i(x,v):$} Here write is performed onto the local memory by transaction $T_i$. This operation appends the data tuple $\langle x, v \rangle$ into the WS of transaction $T_i$.
	
	\vspace{1mm}
	\noindent
	\textit{$\tryc_i():$} This operation is invoked when a transaction $T_i$ has completed all its \op{s} and wants to commit. This operation first checks whether $T_i$ is read-only transaction or not. If it is read-only then it returns commit. Otherwise, for each \tobj{} $x$ (accessed in a predefined order) in $T_i$'s write set, the following check is performed: if timestamp of $T_i$, $i$, lies between the timestamps of the $T_j$ and $T_k$, where transaction $T_k$ reads $x$ from transaction $T_j$, i.e $j < i < k$, then the transaction $T_i$ is aborted. 
	
	If this check succeeds for all the \tobj{s} written by $T_i$, then the version tuples are appended to the version lists and the transaction $T_i$ is committed. Before returning either commit or abort, the transaction id $i$ is removed from \livel.
	
	The system orders all the \tobj{s} ever accessed as $x_1, x_2, ...., x_n$ by any transaction (assuming that the system accesses a total of $n$ \tobj{s}). In this \op, each transaction locks and access \tobj{s} in this increasing order which ensures that the system does not deadlock.
	
	\vspace{1mm}
	\noindent
	\textit{$\checkv(i,x):$} This function checks the version list of $x$. For all version tuples $\langle j, v, rl \rangle$ in $x.vl$ and for all transactions $T_k$ in $rl$, it checks if the timestamp of $T_i$ is between the timestamp of the $T_j$ and $T_k$, i.e $j < i < k$. If so, it returns false else true.
}

\begin{algorithm}[H] 
	\scriptsize
	\label{alg:init} 
	\caption{STM.$\init()$: Invoked at the start of the STM system. Initializes all the \tobj{s} used by the STM system.}
	\begin{algorithmic}[1]
		\makeatletter\setcounter{ALG@line}{233}\makeatother
		\Procedure{STM.$\init()$}{}		
		\State \tcntr{} $\gets$ 1;
		\ForAll {$x$ used by the STM system}
		\State /* $T_0$ is initializing $\tobj$ $x$ with $\langle key, lock, vl, next \rangle$ and add $\langle ts, val, max_r, rl,$ \vn $\rangle$ to $x.vl$*/
		\State add $\langle 0, 0, 0, nil, nil \rangle$ to $x.vl$; \label{lin:t0-init} 
		\EndFor;
		\EndProcedure
	\end{algorithmic}
\end{algorithm}

\begin{algorithm}[H]
	\scriptsize
	\label{alg:begin} 
	\caption{STM.$\begtrans()$: Invoked by a thread to being a new transaction $T_i$}
	\begin{algorithmic}[1]
		\makeatletter\setcounter{ALG@line}{240}\makeatother
		\Procedure{STM.$\begtrans()$}{}
		\State /*Store the latest value of $\tcntr$ in $i$.*/ 
		\State $i \gets \tcntr$; 
		\State $\tcntr \gets \tcntr.get\&Inc()$; 
		\State return $\langle i \rangle$; \label{lin:begfin}
		\EndProcedure
	\end{algorithmic}
\end{algorithm}

\begin{algorithm}[H]
	\scriptsize
	\label{alg:read} 
	\caption{STM.$read_i(x)$: A transaction $T_i$ reads \tobj{} $x$.}
	\begin{algorithmic}[1]
		\makeatletter\setcounter{ALG@line}{246}\makeatother
		\Procedure{STM.$read_i(x)$}{}
		\State /*Check if the \tobj{} $x$ is in 
		$T_i$ local log*/
		\If {($x \in wset_i$)} 
		\State return $\langle wset_i[x].v \rangle$;
		\ElsIf {($x \in rset_i$)} 
		\State return $\langle rset_i[x].v \rangle$;
		\Else
		\State /*\tobj{} $x$ is not in $T_i$ local log*/ 
		\State $x.lock()$;
		\label{lin:rlock}
		\State /*From $x.vls$, identify the right version tuple (or \vtuple).*/ 
		\State $\langle j, v, max_r, rl, $ \vn$\rangle  = \find(i,x)$;	
		\State /*Get the \mr{} timestamps of the $x$*/ 
		\If {(\mr(x) $< i$)} 
		\State \mr(x) = $i$;
		\EndIf			
		\State $x.unlock()$;
		\State Append the \dtuple$\langle x \rangle$ to $rset_i$ of $T_i$ and store \vtuple$\langle j, v, max_r, rl, \vn \rangle$ in $T_i$; /*Maintain the reference of $j$ as closest tuple (or \ctuple)*/
		\State return $\langle v \rangle$; /* return the value as $v$*/
		\EndIf
		\EndProcedure
	\end{algorithmic}
\end{algorithm}

\begin{algorithm}[H]
	\scriptsize
	\label{alg:select}
	\caption{$\find(i,x)$: Finds the tuple $\langle$ j, v, $max_r$, rl, $\vn$$\rangle$ created by the transaction $T_j$ with the largest timestamp smaller than $i$.}
	\begin{algorithmic}[1]
		\makeatletter\setcounter{ALG@line}{266}\makeatother
		\Procedure{$\find(i,x)$}{}
		\State /*Initialize closest tuple (or \ctuple)*/
		\State \ctuple{} = $\langle 0, 0, 0, nil, nil \rangle$;
		\ForAll {$\langle k,v,max_r,rl,\vn{} \rangle \in x.vl$}
		\If {$(k < i)$} 
		\State  \ctuple{} $= \langle k,v, max_r, rl, \vn{} \rangle$; 
		\EndIf;
		\EndFor;
		\State return $\langle \ctuple \rangle$;
		\EndProcedure
	\end{algorithmic}
\end{algorithm}

\begin{algorithm}[H]
	\scriptsize
	\label{alg:write} 
	\caption{STM.$write_i(x,v)$: A transaction $T_i$ writes into local memory.}
	\begin{algorithmic}[1]
		\makeatletter\setcounter{ALG@line}{276}\makeatother
		\Procedure{STM.$write_i(x,v)$}{}
		\If {($x \in wset_i$)} 
		\State Update the \dtuple$\langle x, v \rangle$ to $wset_i$ of $T_i$;
		\Else
		\State Append the \dtuple$\langle x, v \rangle$ to $wset_i$ of $T_i$;
		\EndIf
		\State return $\langle ok \rangle$;
		\EndProcedure
	\end{algorithmic}
\end{algorithm}

\begin{algorithm}[H]  
	\scriptsize
	\label{alg:checkVersion} 
	\caption{$\checkv(i,x)$: Checks the version list; it returns $true$ or $false$.}
	\begin{algorithmic}[1]
		\makeatletter\setcounter{ALG@line}{284}\makeatother
		\Procedure{$\checkv(i,x)$}{}
		\State /*From $x.vls$, identify the right version tuple (or \vtuple).*/ 
		\State $\langle j, v, max_r, rl,$ \vn$\rangle  = \find(i,x)$;	
		\If {$(i < max_r(j))$}   
		\State return $\langle false \rangle$;  
		\EndIf;
		\State /*Maintain the reference of $j$ as closest tuple (or \ctuple)*/
		\State Append the \vtuple$\langle j, v, rl, max_r, \vn \rangle$ in $T_i$;
		\State return $\langle true \rangle$;
		\EndProcedure
	\end{algorithmic}
\end{algorithm}

\begin{algorithm}[H]
	\scriptsize
	\label{alg:tryc} 
	\caption{$\tryc()$: Returns $ok$ on commit else return abort.}
	\begin{algorithmic}[1]
		\makeatletter\setcounter{ALG@line}{294}\makeatother
		\Procedure{$\tryc()$}{}
		\ForAll {\dtuple$\langle x, v \rangle$ in $wset_i$}
		\State /* Lock the \tobj{s} in a predefined order to avoid deadlocks */
		\State $x.lock()$;	
		\If {$(\checkv(i,x) == false)$}
		\State unlock all the variables locked so far;
		\State return $\langle abort(i) \rangle$;
		\EndIf;
		\EndFor;
		\State /* Successfully done with the validation of $wset_i$ and not yet aborted; So the new write versions can be inserted. */
		\ForAll {\dtuple $\langle x,v \rangle $ in $wset_i$}
		\State /*Set the $(x.ts = i)$, $(x.val = v)$ and $(x.max_r = 0)$*/
		\State Insert \vtuple $\langle i, v, 0, nil, nil \rangle$ into $x.vl$ in the increasing order;  \label{lin:ins-vtuple}
		\State /*\cl{$(i)$} contains all the conflicting transactions of $T_i$*/
		\If{($j.rl$ == $nil$)}/*$j$ is the \ctuple{} of $i$*/
		\State /*Add \emph{Largest Timestamp Smaller (or LTS) than $i$} into $T_i$ \cl{}*/
		\State Insert the $j$ into \cl{$(i, j)$};
		\Else
		\ForAll{$T_k$ in $j.rl$} 
		\State /* $T_k$ has already read the version created by $T_j$ */
		\State /*Add \emph{LTS} than $i$ into $T_i$ \cl{}*/
		\State Insert the $k$ into \cl{$(i, k)$};
		\EndFor;
		\EndIf;
		\State /*Add \emph{Smallest Timestamp Larger (or STL) than $i$} into $T_i$ \cl{}*/
		\State Insert the $i$.\vn{} into \cl{$(i, i.\vn)$};
		
		\State $x.unlock()$;	
		\EndFor;
		\ForAll {\dtuple $\langle x \rangle $ in $rset_i$}
		\State $x.lock()$;	
		\State Append $i$ into $j.rl$; /*$j$ is the \ctuple{} of $i$*/
		\State /*\cl{$(i)$} contains all the conflicting transactions of $T_i$*/
		\State Insert the $j$ into \cl{$(i, j)$};
		\State /*Add \emph{STL} than $i$ into $T_i$ \cl{}*/
		\State Insert the $j$.\vn{} into \cl{$(i, j.\vn)$};
		\State $x.unlock()$;	
		\EndFor;
		\State return $\langle ok \rangle$;
		\EndProcedure
	\end{algorithmic}
\end{algorithm}

\begin{algorithm}[H]
	\scriptsize
	\label{alg:getconfl} 
	\caption{\cl{($i, j$)}: Maintain all the conflicting transactions of $T_i$.}
	\begin{algorithmic}[1]
		\makeatletter\setcounter{ALG@line}{333}\makeatother
		\Procedure{\cl{($i, j$)}}{}
		\State $\cl(i).lock$;
		\State Appending the transaction $j$ into \cl{} of $i$;
		\State $\cl(i).unlock$;	
		\EndProcedure
	\end{algorithmic}
\end{algorithm}

\begin{algorithm}[H]
	\scriptsize
	\label{alg:getconfl} 
	\caption{STM.\gconfl{($i$)}: Get the conflict list corresponding to transaction $T_i$}
	\begin{algorithmic}[1]
		\makeatletter\setcounter{ALG@line}{338}\makeatother
		\Procedure{STM.\gconfl{($i$)}}{}
		\State Transaction $T_i$ search the respective conflict in conflict list (or \cl{}).
		\State return $\langle \cl$(i)$ \rangle$;
		\EndProcedure
	\end{algorithmic}
\end{algorithm}

\begin{algorithm}[H]
	\scriptsize
	\label{alg:ap-abort} 
	\caption{$abort(i)$: Invoked by various STM methods to abort transaction $T_i$. It returns $\mathcal{A}$.}
	\begin{algorithmic}[1]
		\makeatletter\setcounter{ALG@line}{342}\makeatother
		\Procedure{$abort(i)$}{}
		\State unlock all \tobj{s} locked by $T_i$ and release its local log;      
		\State return $\langle \mathcal{A} \rangle$;
		\EndProcedure
	\end{algorithmic}
\end{algorithm}

\input{correctness}

%% file: figs/mvtofun.pdf_t
\begin{picture}(0,0)%
\includegraphics{figs/mvtofun.pdf}%
\end{picture}%
\setlength{\unitlength}{4144sp}%
\begingroup\makeatletter\ifx\SetFigFont\undefined%
\gdef\SetFigFont#1#2#3#4#5{%
  \reset@font\fontsize{#1}{#2pt}%
  \fontfamily{#3}\fontseries{#4}\fontshape{#5}%
  \selectfont}%
\fi\endgroup%
\begin{picture}(10709,9797)(2724,-15371)
\put(2881,-6361){\makebox(0,0)[lb]{\smash{{\SetFigFont{17}{20.4}{\rmdefault}{\mddefault}{\updefault}{\color[rgb]{0,0,0}$x$}%
}}}}
\put(2791,-5911){\makebox(0,0)[lb]{\smash{{\SetFigFont{17}{20.4}{\rmdefault}{\mddefault}{\updefault}{\color[rgb]{0,0,0}$key$}%
}}}}
\put(3286,-5911){\makebox(0,0)[lb]{\smash{{\SetFigFont{17}{20.4}{\rmdefault}{\mddefault}{\updefault}{\color[rgb]{0,0,0}$Lock$}%
}}}}
\put(4411,-5911){\makebox(0,0)[lb]{\smash{{\SetFigFont{17}{20.4}{\rmdefault}{\mddefault}{\updefault}{\color[rgb]{0,0,0}$next$}%
}}}}
\put(4006,-5911){\makebox(0,0)[lb]{\smash{{\SetFigFont{17}{20.4}{\rmdefault}{\mddefault}{\updefault}{\color[rgb]{0,0,0}$vl$}%
}}}}
\put(9721,-8521){\makebox(0,0)[lb]{\smash{{\SetFigFont{17}{20.4}{\rmdefault}{\mddefault}{\updefault}{\color[rgb]{1,0,0}7}%
}}}}
\put(10801,-14596){\makebox(0,0)[lb]{\smash{{\SetFigFont{17}{20.4}{\rmdefault}{\mddefault}{\updefault}{\color[rgb]{1,0,0}$T_{10}$}%
}}}}
\put(12691,-14641){\makebox(0,0)[lb]{\smash{{\SetFigFont{17}{20.4}{\rmdefault}{\mddefault}{\updefault}{\color[rgb]{0,.56,0}$T_5$}%
}}}}
\put(4501,-7216){\makebox(0,0)[lb]{\smash{{\SetFigFont{17}{20.4}{\rmdefault}{\mddefault}{\updefault}{\color[rgb]{0,0,0}$ts$}%
}}}}
\put(6346,-7216){\makebox(0,0)[lb]{\smash{{\SetFigFont{17}{20.4}{\rmdefault}{\mddefault}{\updefault}{\color[rgb]{0,0,0}$vNext$}%
}}}}
\put(6031,-7216){\makebox(0,0)[lb]{\smash{{\SetFigFont{17}{20.4}{\rmdefault}{\mddefault}{\updefault}{\color[rgb]{0,0,0}$rl$}%
}}}}
\put(4861,-7216){\makebox(0,0)[lb]{\smash{{\SetFigFont{17}{20.4}{\rmdefault}{\mddefault}{\updefault}{\color[rgb]{0,0,0}$val$}%
}}}}
\put(5311,-7216){\makebox(0,0)[lb]{\smash{{\SetFigFont{17}{20.4}{\rmdefault}{\mddefault}{\updefault}{\color[rgb]{0,0,0}$max_r$}%
}}}}
\put(4546,-7666){\makebox(0,0)[lb]{\smash{{\SetFigFont{17}{20.4}{\rmdefault}{\mddefault}{\updefault}{\color[rgb]{0,0,0}0}%
}}}}
\put(4951,-7666){\makebox(0,0)[lb]{\smash{{\SetFigFont{17}{20.4}{\rmdefault}{\mddefault}{\updefault}{\color[rgb]{0,0,0}$v_0$}%
}}}}
\put(5491,-7666){\makebox(0,0)[lb]{\smash{{\SetFigFont{17}{20.4}{\rmdefault}{\mddefault}{\updefault}{\color[rgb]{0,0,0}0}%
}}}}
\put(5941,-7666){\makebox(0,0)[lb]{\smash{{\SetFigFont{17}{20.4}{\rmdefault}{\mddefault}{\updefault}{\color[rgb]{0,0,0}$nil$}%
}}}}
\put(10801,-7216){\makebox(0,0)[lb]{\smash{{\SetFigFont{17}{20.4}{\rmdefault}{\mddefault}{\updefault}{\color[rgb]{0,.56,0}$ts$}%
}}}}
\put(12646,-7216){\makebox(0,0)[lb]{\smash{{\SetFigFont{17}{20.4}{\rmdefault}{\mddefault}{\updefault}{\color[rgb]{0,.56,0}$vNext$}%
}}}}
\put(12331,-7216){\makebox(0,0)[lb]{\smash{{\SetFigFont{17}{20.4}{\rmdefault}{\mddefault}{\updefault}{\color[rgb]{0,.56,0}$rl$}%
}}}}
\put(11161,-7216){\makebox(0,0)[lb]{\smash{{\SetFigFont{17}{20.4}{\rmdefault}{\mddefault}{\updefault}{\color[rgb]{0,.56,0}$val$}%
}}}}
\put(11611,-7216){\makebox(0,0)[lb]{\smash{{\SetFigFont{17}{20.4}{\rmdefault}{\mddefault}{\updefault}{\color[rgb]{0,.56,0}$max_r$}%
}}}}
\put(7651,-7216){\makebox(0,0)[lb]{\smash{{\SetFigFont{17}{20.4}{\rmdefault}{\mddefault}{\updefault}{\color[rgb]{.56,0,.56}$ts$}%
}}}}
\put(9496,-7216){\makebox(0,0)[lb]{\smash{{\SetFigFont{17}{20.4}{\rmdefault}{\mddefault}{\updefault}{\color[rgb]{.56,0,.56}$vNext$}%
}}}}
\put(9181,-7216){\makebox(0,0)[lb]{\smash{{\SetFigFont{17}{20.4}{\rmdefault}{\mddefault}{\updefault}{\color[rgb]{.56,0,.56}$rl$}%
}}}}
\put(8011,-7216){\makebox(0,0)[lb]{\smash{{\SetFigFont{17}{20.4}{\rmdefault}{\mddefault}{\updefault}{\color[rgb]{.56,0,.56}$val$}%
}}}}
\put(8461,-7216){\makebox(0,0)[lb]{\smash{{\SetFigFont{17}{20.4}{\rmdefault}{\mddefault}{\updefault}{\color[rgb]{.56,0,.56}$max_r$}%
}}}}
\put(7696,-7666){\makebox(0,0)[lb]{\smash{{\SetFigFont{17}{20.4}{\rmdefault}{\mddefault}{\updefault}{\color[rgb]{.56,0,.56}5}%
}}}}
\put(8686,-7666){\makebox(0,0)[lb]{\smash{{\SetFigFont{17}{20.4}{\rmdefault}{\mddefault}{\updefault}{\color[rgb]{.56,0,.56}5}%
}}}}
\put(8101,-7666){\makebox(0,0)[lb]{\smash{{\SetFigFont{17}{20.4}{\rmdefault}{\mddefault}{\updefault}{\color[rgb]{.56,0,.56}$v_5$}%
}}}}
\put(12736,-7666){\makebox(0,0)[lb]{\smash{{\SetFigFont{17}{20.4}{\rmdefault}{\mddefault}{\updefault}{\color[rgb]{0,.56,0}$nil$}%
}}}}
\put(10756,-7666){\makebox(0,0)[lb]{\smash{{\SetFigFont{17}{20.4}{\rmdefault}{\mddefault}{\updefault}{\color[rgb]{0,.56,0}10}%
}}}}
\put(11746,-7666){\makebox(0,0)[lb]{\smash{{\SetFigFont{17}{20.4}{\rmdefault}{\mddefault}{\updefault}{\color[rgb]{0,.56,0}10}%
}}}}
\put(12241,-7666){\makebox(0,0)[lb]{\smash{{\SetFigFont{17}{20.4}{\rmdefault}{\mddefault}{\updefault}{\color[rgb]{0,.56,0}$nil$}%
}}}}
\put(7966,-6361){\makebox(0,0)[lb]{\smash{{\SetFigFont{20}{24.0}{\rmdefault}{\mddefault}{\updefault}{\color[rgb]{.75,.38,0}$Version$ $List$ $(or$ $vl)$}%
}}}}
\put(2746,-9781){\makebox(0,0)[lb]{\smash{{\SetFigFont{17}{20.4}{\rmdefault}{\mddefault}{\updefault}{\color[rgb]{0,0,0}$T_0$}%
}}}}
\put(3646,-9511){\makebox(0,0)[lb]{\smash{{\SetFigFont{17}{20.4}{\rmdefault}{\mddefault}{\updefault}{\color[rgb]{0,0,0}$w_0(x)$}%
}}}}
\put(4816,-9511){\makebox(0,0)[lb]{\smash{{\SetFigFont{17}{20.4}{\rmdefault}{\mddefault}{\updefault}{\color[rgb]{0,0,0}$C_0$}%
}}}}
\put(8731,-10096){\makebox(0,0)[lb]{\smash{{\SetFigFont{17}{20.4}{\rmdefault}{\mddefault}{\updefault}{\color[rgb]{.56,0,.56}$C_5$}%
}}}}
\put(11881,-10681){\makebox(0,0)[lb]{\smash{{\SetFigFont{17}{20.4}{\rmdefault}{\mddefault}{\updefault}{\color[rgb]{1,0,0}$C_7$}%
}}}}
\put(6751,-10186){\makebox(0,0)[lb]{\smash{{\SetFigFont{17}{20.4}{\rmdefault}{\mddefault}{\updefault}{\color[rgb]{.56,0,.56}$w_5(x)$}%
}}}}
\put(8866,-11401){\makebox(0,0)[lb]{\smash{{\SetFigFont{17}{20.4}{\rmdefault}{\mddefault}{\updefault}{\color[rgb]{0,.56,0}$w_{10}(x)$}%
}}}}
\put(10576,-10726){\makebox(0,0)[lb]{\smash{{\SetFigFont{17}{20.4}{\rmdefault}{\mddefault}{\updefault}{\color[rgb]{1,0,0}$r_7(x)$}%
}}}}
\put(5986,-11131){\makebox(0,0)[lb]{\smash{{\SetFigFont{17}{20.4}{\rmdefault}{\mddefault}{\updefault}{\color[rgb]{1,0,0}$T_7$}%
}}}}
\put(6751,-11716){\makebox(0,0)[lb]{\smash{{\SetFigFont{17}{20.4}{\rmdefault}{\mddefault}{\updefault}{\color[rgb]{0,.56,0}$T_{10}$}%
}}}}
\put(10351,-11716){\makebox(0,0)[lb]{\smash{{\SetFigFont{17}{20.4}{\rmdefault}{\mddefault}{\updefault}{\color[rgb]{0,.56,0}$C_{10}$}%
}}}}
\put(4951,-10501){\makebox(0,0)[lb]{\smash{{\SetFigFont{17}{20.4}{\rmdefault}{\mddefault}{\updefault}{\color[rgb]{.56,0,.56}$T_5$}%
}}}}
\put(11161,-7666){\makebox(0,0)[lb]{\smash{{\SetFigFont{17}{20.4}{\rmdefault}{\mddefault}{\updefault}{\color[rgb]{0,.56,0}$v_{10}$}%
}}}}
\put(5581,-13291){\makebox(0,0)[lb]{\smash{{\SetFigFont{17}{20.4}{\rmdefault}{\mddefault}{\updefault}{\color[rgb]{.56,0,.56}$ts$}%
}}}}
\put(6661,-13291){\makebox(0,0)[lb]{\smash{{\SetFigFont{17}{20.4}{\rmdefault}{\mddefault}{\updefault}{\color[rgb]{.56,0,.56}$cl$}%
}}}}
\put(6976,-13291){\makebox(0,0)[lb]{\smash{{\SetFigFont{17}{20.4}{\rmdefault}{\mddefault}{\updefault}{\color[rgb]{.56,0,.56}$cNext$}%
}}}}
\put(5581,-13741){\makebox(0,0)[lb]{\smash{{\SetFigFont{17}{20.4}{\rmdefault}{\mddefault}{\updefault}{\color[rgb]{.56,0,.56}$T_5$}%
}}}}
\put(8326,-13291){\makebox(0,0)[lb]{\smash{{\SetFigFont{17}{20.4}{\rmdefault}{\mddefault}{\updefault}{\color[rgb]{1,0,0}$ts$}%
}}}}
\put(8731,-13291){\makebox(0,0)[lb]{\smash{{\SetFigFont{17}{20.4}{\rmdefault}{\mddefault}{\updefault}{\color[rgb]{1,0,0}$Lock$}%
}}}}
\put(9406,-13291){\makebox(0,0)[lb]{\smash{{\SetFigFont{17}{20.4}{\rmdefault}{\mddefault}{\updefault}{\color[rgb]{1,0,0}$cl$}%
}}}}
\put(9721,-13291){\makebox(0,0)[lb]{\smash{{\SetFigFont{17}{20.4}{\rmdefault}{\mddefault}{\updefault}{\color[rgb]{1,0,0}$cNext$}%
}}}}
\put(8326,-13741){\makebox(0,0)[lb]{\smash{{\SetFigFont{17}{20.4}{\rmdefault}{\mddefault}{\updefault}{\color[rgb]{1,0,0}$T_7$}%
}}}}
\put(11071,-13291){\makebox(0,0)[lb]{\smash{{\SetFigFont{17}{20.4}{\rmdefault}{\mddefault}{\updefault}{\color[rgb]{0,.56,0}$ts$}%
}}}}
\put(11476,-13291){\makebox(0,0)[lb]{\smash{{\SetFigFont{17}{20.4}{\rmdefault}{\mddefault}{\updefault}{\color[rgb]{0,.56,0}$Lock$}%
}}}}
\put(12151,-13291){\makebox(0,0)[lb]{\smash{{\SetFigFont{17}{20.4}{\rmdefault}{\mddefault}{\updefault}{\color[rgb]{0,.56,0}$cl$}%
}}}}
\put(12466,-13291){\makebox(0,0)[lb]{\smash{{\SetFigFont{17}{20.4}{\rmdefault}{\mddefault}{\updefault}{\color[rgb]{0,.56,0}$cNext$}%
}}}}
\put(3826,-13741){\makebox(0,0)[lb]{\smash{{\SetFigFont{17}{20.4}{\rmdefault}{\mddefault}{\updefault}{\color[rgb]{0,0,0}$nil$}%
}}}}
\put(2836,-13291){\makebox(0,0)[lb]{\smash{{\SetFigFont{17}{20.4}{\rmdefault}{\mddefault}{\updefault}{\color[rgb]{0,0,0}$ts$}%
}}}}
\put(3241,-13291){\makebox(0,0)[lb]{\smash{{\SetFigFont{17}{20.4}{\rmdefault}{\mddefault}{\updefault}{\color[rgb]{0,0,0}$Lock$}%
}}}}
\put(3916,-13291){\makebox(0,0)[lb]{\smash{{\SetFigFont{17}{20.4}{\rmdefault}{\mddefault}{\updefault}{\color[rgb]{0,0,0}$cl$}%
}}}}
\put(4231,-13291){\makebox(0,0)[lb]{\smash{{\SetFigFont{17}{20.4}{\rmdefault}{\mddefault}{\updefault}{\color[rgb]{0,0,0}$cNext$}%
}}}}
\put(2836,-13741){\makebox(0,0)[lb]{\smash{{\SetFigFont{17}{20.4}{\rmdefault}{\mddefault}{\updefault}{\color[rgb]{0,0,0}$T_0$}%
}}}}
\put(12646,-13741){\makebox(0,0)[lb]{\smash{{\SetFigFont{17}{20.4}{\rmdefault}{\mddefault}{\updefault}{\color[rgb]{0,.56,0}$nil$}%
}}}}
\put(5986,-13291){\makebox(0,0)[lb]{\smash{{\SetFigFont{17}{20.4}{\rmdefault}{\mddefault}{\updefault}{\color[rgb]{.56,0,.56}$Lock$}%
}}}}
\put(7201,-14641){\makebox(0,0)[lb]{\smash{{\SetFigFont{17}{20.4}{\rmdefault}{\mddefault}{\updefault}{\color[rgb]{.56,0,.56}$T_0$}%
}}}}
\put(9946,-14596){\makebox(0,0)[lb]{\smash{{\SetFigFont{17}{20.4}{\rmdefault}{\mddefault}{\updefault}{\color[rgb]{1,0,0}$T_5$}%
}}}}
\put(11071,-13741){\makebox(0,0)[lb]{\smash{{\SetFigFont{17}{20.4}{\rmdefault}{\mddefault}{\updefault}{\color[rgb]{0,.56,0}$T_{10}$}%
}}}}
\put(7291,-8971){\makebox(0,0)[lb]{\smash{{\SetFigFont{20}{24.0}{\rmdefault}{\mddefault}{\updefault}{\color[rgb]{0,0,0}(a) Version List}%
}}}}
\put(7381,-15271){\makebox(0,0)[lb]{\smash{{\SetFigFont{20}{24.0}{\rmdefault}{\mddefault}{\updefault}{\color[rgb]{0,0,0}(c) Conflict List}%
}}}}
\put(6391,-12526){\makebox(0,0)[lb]{\smash{{\SetFigFont{20}{24.0}{\rmdefault}{\mddefault}{\updefault}{\color[rgb]{0,0,0}(b) Transction Timeline View}%
}}}}
\end{picture}%

%% file: correctness.tex
\section{Correctness}
\label{sec:correctness}

\subsection{The Linearization Points of Lock-free Graph Library Methods}

Here, we list the linearization points (LPs) of each method as follows:

\begin{enumerate}
	\item \addv{(\vgn)}: (\vp.\vn.CAS(\vc, \vgn)) in \Lineref{addv5} is the LP point of \addv{()} method if \vnode{} is not exist in the BG. If \vnode{} is exist in the BG then (\vc.$ts_i$ $\neq$ \vgn.$ts_i$) in \Lineref{addv3} is the LP point.
	
	\item \adde{\emph{(fromNode, toNode)}}: 		(\ep.\en.CAS(\ec, \egn)) in \Lineref{adde7} is the LP point of \adde{()} method if \enode{} is not exist in the BG. If \enode{} is exist in the BG then 		(\ec.$ts_i$ $\neq$ toNode.$ts_i$) in \Lineref{adde5} is the LP point.
	
	\item \searchl{(cacheVer, $AU_{id}$)}: (cacheVer.\inc.CAS(0, -1)) in \Lineref{sl1} is the LP point of \searchl{()} method.
	
	\item \searchg{(BG, $AU_{id}$)}: (\vnode.\inc.CAS(0, -1)) in \Lineref{sg1} is the LP point of \searchg{()} method.
	
	\item \emph{decInCount}(remNode): \Lineref{ren1} is the LP point of \emph{decInCount()} method.
\end{enumerate}

\begin{theorem}
	All the dependencies between the conflicting nodes are captured in the BG.
\end{theorem} 

\begin{proof}
	Dependencies between the conflicting nodes are captured in the BG with the help of LP points of lock-free graph library methods defined above.
\end{proof}